\documentclass[pss]{wiley2sp} 
\usepackage{amsmath}
\usepackage{bm}              
\usepackage{hyperref}

\begin{document}

\title{Interplay of Rashba/Dresselhaus spin splittings probed by photogalvanic spectroscopy (review)
}

\titlerunning{Interplay of Rashba/Dresselhaus spin splittings}

\author{%
  Sergey D. Ganichev \textsuperscript{\Ast,\textsf{\bfseries 1}},
  Leonid E. Golub    \textsuperscript{\textsf{\bfseries 2}},
    }

\authorrunning{S.D. Ganichev and L.E. Golub}

\mail{e-mail
  \textsf{sergey.ganichev@ur.de}, Phone:
  +49-941-9432050, Fax: +49-941-9431657}

\institute{%
  \textsuperscript{1}\,University of Regensburg, 93040 Regensburg, Germany \\
  \textsuperscript{2}\,Ioffe Physical-Technical Institute of the RAS, 194021 St. Petersburg, Russia
}

\received{XXXX, revised XXXX, accepted XXXX} 
\published{XXXX} 


\abstract{
%
%
%
%
\abstcol{%

The paper reviews the interplay of Rashba/Dresselhaus spin 
splittings in various two dimensional systems made of III-V, wurtzite and SiGe. 
We discuss the  symmetry aspects of the linear and cubic in 
electron wavevector spin splitting in heterostructures 
prepared on (001)-, (110)-, (111)-, (113)-, (112)-, and (013)-
oriented substrates and address the requirements for suppression 
of spin}%
{relaxation and realization of the persistent spin helix state.
In experimental part of the paper we overview 
experimental results on the interplay of Rashba/Dresselhaus 
spin splittings probed by photogalvanic spectroscopy: 
the method based on the phenomenological equivalence 
of the linear-in-wavevector spin splitting and 
several photogalvanic phenomena. 
}
}
%
%

\maketitle   

\section{Introduction}
\label{intro}

Quantum phenomena in semiconductors  are highly sensitive to
subtle details of the carrier energy spectrum so that even a small
spin splitting of energy  bands  may result in measurable effects.
A textbook example of the band spin splitting is the Zeeman effect,
which is caused by the coupling of an external magnetic field and electron spin. 
However, band spin degeneracy can also be removed without action of a magnetic field. 
This phenomenon is caused by the spin-orbit interaction (SOI) in 
non-centrosymmetric  crystals, a relativistic effect 
allowing for coupling of electron spin and orbital degrees of freedom.
As a result the spin degeneracy of the energy bands is lifted even 
in nonmagnetic materials.
This coupling is described by a Hamiltonian with products of $\bm \sigma$ and $\bm k$ terms where
$\bm \sigma$ are the Pauli spin matrices and $\bm k$ is the electron wave vector. 
The origin of these terms are the bulk inversion asymmetry (BIA) and  
the structure inversion asymmetry (SIA).
Microscopically, BIA  stems from the absence of the inversion 
symmetry in the bulk material and gives rise to the Dresselhaus 
spin splitting  in  bulk  and 
low-dimensional semiconductors~\cite{A1Dresselhaus55p580,A1Dyakonov86p110}. 
By contrast, SIA originates from the inversion asymmetry of the 
confining potential and yields the Rashba term in the Hamiltonian
whose strength can be manipulated by an external field (Rashba effect)~\cite{A1Rashba60p1109,A1Bychkov84p78}.
In particular, the SIA/BIA coupling of the electron wavevector and spin
causes a Larmor precession in an internal $\bm k$-dependent  effective magnetic field
for electrons moving through a semiconductor structure.
Note that in addition to BIA and SIA an interface inversion asymmetry (IIA) may
yield {\boldmath$k$}-linear terms caused by non-inversion symmetric
bonding of atoms at heterostructure
interfaces~\cite{Krebs1996,Krebs1997,A1Vervoort97p12744}.
Since the IIA results in the spin-orbit interaction of 
the same form as BIA in (001)-grown III-V systems, 
we  disregard this type of spin splitting  in further consideration.

While the existence of the zero-magnetic field spin splitting 
is known since the sixties of the last century~\cite{A1Dresselhaus55p580,A1Rashba60p1109}
it quickens an enormous interest 
since manipulation of the electron 
spin instead of its charge 
has been considered as a candidate 
for the  future electronics -- 
spintronics. 
The cause  of this interest is that the 
Rashba effect in  two-dimensional electron systems (2DES) 
provides a unique possibility to manipulate electron spin by
means of external electric field and is 
of great importance for the generation, manipulation 
and detection of spin currents as well as for control of the 
spin relaxation processes in low dimensional 
semiconductors, for  reviews 
see~\cite{spintronicbook02,Winkler2003,Maekawa,Fabian08,Dyakonovbook,Awschalombook2008,Awschalombook2009,Awschalombook2010,Wu10,HandbookZutic,SemicSpintronics,SpinCurrent}.
In particular, spin manipulation by
means of electric field, pure spin currents and electric currents
caused by spin polarization have attracted continuously growing  interest
from both the experimental and theoretical points of view. 
Most of these works are aimed to two-dimensional systems,  where BIA and SIA terms couple 
the \textit{in-plane} wavevector of confined  electrons $\bm k$ with the 
\textit{in-} or \textit{out-of-plane} components of the electron spin $\bm S$. 
The relative orientation of the coupled $\bm k$- and  
$\bm S$-components is determined by the symmetry of
the system. Consequently, it depends on the QW growth plane crystallographic orientation and 
on the considered direction of the in-plane wavevector. In many 
systems the SIA and BIA terms can interfere resulting in an anisotropy
of the spin splitting. The strongest anisotropy can be achieved 
in (001)-grown quantum wells (QWs) with the $\bm k$-linear 
Rashba and Dresselhaus terms  of equal strength.
Under these circumstances,  the dominant mechanism of spin
dephasing (Dyakonov-Perel  relaxation~\cite{DP71}) 
is suppressed~\cite{Averkiev99PRB15582,Averkiev02pR271,omega4}
making possible a diffusive spin field transistor~\cite{Schliemann03p146801} as well as
giving rise to a persistent spin helix predicted in Ref.~\cite{Bernevig2006} 
and observed in GaAs low dimensional systems~\cite{Koralek2009,Walser2012Nature}.
In fact, for this particular case, the spin splitting  vanishes in
certain {$\bm k$}-space directions and an effective magnetic
field caused by SOI is aligned along a certain crystallographic 
axis for all $\bm k$ being ineffective for spins oriented 
along this axis (see, e.g. Ref.~\cite{Averkiev02pR271,Schliemann03p146801,review2003spin}). 
Further important example of the SIA/BIA anisotropy 
is manipulation of the spin dephasing 
in quantum wells grown on (110) or (111) crystallographic 
planes  where extremely long spin relaxation times have been 
experimentally achieved
by adjusting of  Rashba and Dresselhaus spin splitting~\cite{Ohno1999,Harley03,Dohrmann2004,Hall2005,PRL08110,Roemer2007,Mueller2008,Mueller10,Balocchi2011,Griesbeck2012,Ye2012,Biermann2012,Hernandez2012,Wang2013}. 
Lately, there has been much effort in the studying of the SIA/BIA-interplay both
theoretically with new device proposals~\cite{Schliemann03p146801,Cartoixa2003,Hall2003APL,Cartoixa2005,JapaneseworkAPL2012}
and experimentally with the aim to obtain 
particular relation between SIA and BIA spin splitting
in QW systems of various crystallographic 
orientations.

Owing to the fact that Rashba/Dresselhaus zero magnetic field spin splittings give 
rise to a large number of diverse physical phenomena their characterization and control 
are of fundamental importance for spin physics in semiconductors. 
The relative orientation of spin and electron wavevector in eigenstates
and strength of these splittings  
depend on macroscopic conditions such as 
structure crystallographic orientation, QW width, temperature,  
electron density, doping profile, stress, etc.
Consequently, the interplay of the Rashba/Dresselhaus spin splittings 
is strongly affected by these parameters and requires a detailed study. 
Various methods  providing an experimental access to the SIA/BIA interplay have been developed to which belong 
i) investigations of the anisotropy of the Raman effect~\cite{Jusserand92,Jusserand95p4707};
ii) study of the weak antilocalization (WAL)~\cite{Pikus95p16928,Knap96,Miller03,Glazov2009,Yu2008} and
WAL in tilted magnetic fields~\cite{Minkov2004,Scheid2007,Scheid2009,Kunihashi09}; 
iii) photogalvanic effects~\cite{PRL08110,PRL04,PRB07,LechnerAPL2009,Bieler05,Yang06p186605,Tang2007GaN,Zhao07GaN,Frazier2009KhodropPGE,Yu2012_1};
iv) investigation of spin-relaxation anisotropy by Hanle-effect~\cite{GolubKochereshko};
v) studying of the gate dependence of spin relaxation~\cite{Eldridge08,Larionov2008,Eldridge2010_2,Eldridge2011}, 
as well as 
vi) experiments on time resolved Kerr effect or Faraday rotation in 
special experimental 
geometries~\cite{Stich07,Cheng08,Stich07_2,Meier07,Korn08,Meier08,Studer09,Studer10}, 
including 
magnetooptical Kerr effect with in-plane magnetic fields~\cite{Stich07_2} 
and optical monitoring the angular dependence
of the electron spin precession on their direction of motion with respect to the crystal lattice~\cite{Meier07}.
The multifaceted  SIA/BIA spin splitting has been the subject of 
a tremendous number of works and numerous reviews. 
Our contribution to this special issue 
is primary focused on the results obtained in the framework 
within the DFG Schwerpunktprogramm SPP 1285 and, consequently, limited 
to the investigation of SIA/BIA explored by study of the  photogalvanic 
effects anisotropy. The developed methods are based on the phenomenological 
equivalence of SIA/BIA spin splitting and several photogalvanic 
phenomena~\cite{review2003spin,IvchenkoGanichev,GanichevPrettl}, 
which all have a common property: They are described by the linear 
coupling of a polar vector and an axial vector,
like  the  electron wavevector with its spin 
in Rashba/Dresselhaus effect or, e.g., electric current 
with an average non-equilibrium spin in 
the spin-galvanic effect~\cite{PRL04}. 
Indeed, such phenomena are
described by second rank pseudo-tensors whose irreducible
components differ by a scalar factor only. 
Therefore, these methods allow determination of 
the spin-orbit coupling anisotropy in 2DES
and do not require a knowledge of microscopic 
details or rely on theoretical quantities.
Furthermore, previous studies demonstrated that
the discussed effects are very general and 
measurable signals can be obtained for almost 
all 2DES and even  at room temperature (for reviews see e.g. Ref.~\cite{IvchenkoGanichev}).
Thus photogalvanic experiments allow 
characterization of the SIA/BIA interplay 
upon variation of macroscopic parameters in 
a wide range.

The paper is organized in the following way:
in section~\ref{symmetry} an overview of
the  symmetry aspects of the Rashba/Dresselhaus effects 
in   III-V semiconductor materials  is given. 
First the removal of spin degeneracy due to spin-orbit
interaction is addressed and then the SIA/BIA spin
splitting in {\boldmath$k$}-space for 2DES 
grown in various crystallographic directions is presented.
Sections~\ref{method} and \ref{technique} introduce the method based on photogalvanics and 
give a short account for the experimental
technique, respectively. 
The experimental results on interplay of SIA/BIA upon variation of  
2DES design and characteristics are presented and discussed 
in sections~\ref{IIIV} (III-V-based QWs),~\ref{wurtziteexp} (wurtzite 2DES) and~\ref{SiGEexp} (SiGe QWs). Conclusions and outlook are given in section~\ref{conclusions}.

\begin{table}
    \centering
    \label{table1}
    \caption{       Correspondence between growth-orientation dependent 
$x,y,z$ labels and crystallografic orientations. Note that in (001)-grown III-V material-based 
QWs  in a valuable number of works aimed to SIA/BIA spin splitting
cubic axes with $x'\parallel$~[100] and  $y'\parallel$~[010] are used.}
        \begin{tabular}{cccccc|c}
\hline \hline 
    \multicolumn{7}{c}{Growth   plane} \\ \hline
    \multicolumn{6}{c|}{III-V and SiGe}  & Wurtzite \\ \hline
    & bulk  & (001) & (110) & (111) &(113)&  \\ \hline
$x$ &[100] & $[1\bar{1}0]$ & $[\bar{1}10]$ & $[11\bar{2}]$ &$[1\bar{1}0]$& $[11\bar{2}0]$\\ 
$y$ &[010] & [110] & $[00{1}]$ & $[\bar{1}10]$ &$[33\bar{2}]$& $[1\bar{1}00]$\\ 
$z$ &[001] & [001] & $[110]$ & $[111]$ & [113]& [0001]\\ \hline \hline
        \end{tabular}
\end{table}

\section{Symmetry analysis of the Rashba/Dresselhaus band spin splitting in III-V materials}
\label{symmetry}

In the absence of external magnetic fields the time inversion 
results in the Kramers theorem which reads as $\varepsilon_{\uparrow}(\bm k) = \varepsilon_{\downarrow}(-\bm k)$. 
Here $\varepsilon$ is electron energy, and $\uparrow$/$\downarrow$ enumerate two spin states. 
If the system has an inversion center then, applying the space 
inversion operation, one gets $\varepsilon_{\uparrow}(\bm k) = \varepsilon_{\uparrow}(-\bm k)$. 
Combining these two results we see that two spin states with the same 
wavevector $\bm k$ have the same energy
and
the electron energy spectrum in the conduction band minima is well described by a 
parabolic  dispersion: 
$\varepsilon_{\uparrow}(\bm k) = \varepsilon_{\downarrow}(\bm k) = \hbar^2k^2/(2m^*)$, 
where $m^*$ is the effective mass in the conduction band. 

However, in the system lacking an inversion center, e.g.  III-V and wurtzite bulk semiconductors and 2DES,
the spin splitting can be present even in zero magnetic field. 
Such a splitting is caused by  spin-orbit interaction. 
The corresponding Hamiltonian $H_{\rm SO}$ is given by a sum of 
products of the Pauli matrices and odd  combinations of the wavevector components. 
In bulk III-V semiconductors belonging to  T$_d$ 
point group symmetry  it is described by the 
cubic in the 3D wavevector $\bm k$ terms 
introduced by Dresselhaus~\cite{A1Dresselhaus55p580}:
\begin{equation}
\label{Dress_bulk}
    H_{\rm bulk} = \gamma [\sigma_xk_x(k_y^2-k_z^2)+\sigma_yk_y(k_z^2-k_x^2)+\sigma_zk_z(k_x^2-k_y^2)].
\end{equation}
Here $\gamma$ is the only one linearly-independent constant for the $T_d$ point group
and $x,y,z$ are cubic axes. Note that hereafter the crystallographic orientation of  $x,y,z$ axes 
for each considered system is given in 
Table~\ref{table1}. Despite this splitting determines the 
Dyakonov-Perel spin relaxation rate, its value can not be manipulated by an electric field
and  is determined by the constant $\gamma$.

\begin{figure}[tb]
\begin{center}
\includegraphics[width=\linewidth]{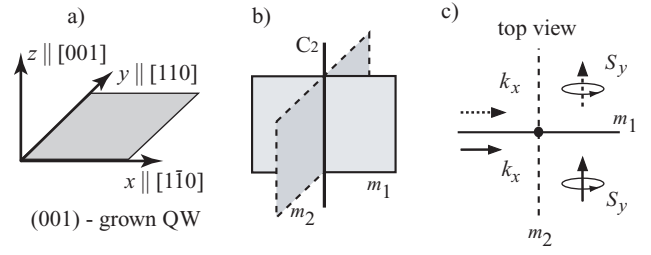}
\end{center}
\caption{(a) Coordinate system used for (001)-grown III-V QW, 
(b)  symmetry elements of the C$_{2\mathrm{v}}$ point group: 
mirror planes $m_1$ and  $m_2$ and C$_2$-axis in the QW grown along $z\,
\parallel [001]$. 
Arrows in the drawing (c) show  that the
reflection in the mirror plane $m_1$ does not change the sign of both the
polar vector component $k_x$ and the axial vector component $S_y$, 
demonstrating that a linear coupling $k_x$ and $S_y$ is 
allowed under this symmetry operation. This
coupling is also allowed by the other symmetry 
operations (mirror reflection by the plane $m_2$ at which both components change they sign
and the C$_2$-axis) rotation of the point group 
yielding the $k_x \sigma_y$ 
terms in the effective Hamiltonian.}
\label{planes001}
\end{figure}

In 2D systems, confinement and symmetry lowering result in a 
more rich spin-orbit interaction, which is described by 
new terms in the Hamiltonian both, linear and cubic, 
in the electron 2D wavevector. The corresponding spin-orbit splitting is sensitive to 
external parameters like electric field, temperature, structure design, 
crystallographic orientation  etc. Below we consider one by one QW 
structures grown in various directions. The three point groups 
D$_{2d}$, C$_{2v}$ and C$_s$ are particularly relevant for zinc-blende structure 
based QWs~\cite{review2003spin,A1book,A1book2}. Hereafter the
Sch{\"o}nflies notation is used to label the point groups. In the
international notation they are labeled as $\bar{4}2m$, $mm2$ and
$m$, respectively.

\begin{figure*}[tb]
\begin{center}
\includegraphics[width=0.8\linewidth]{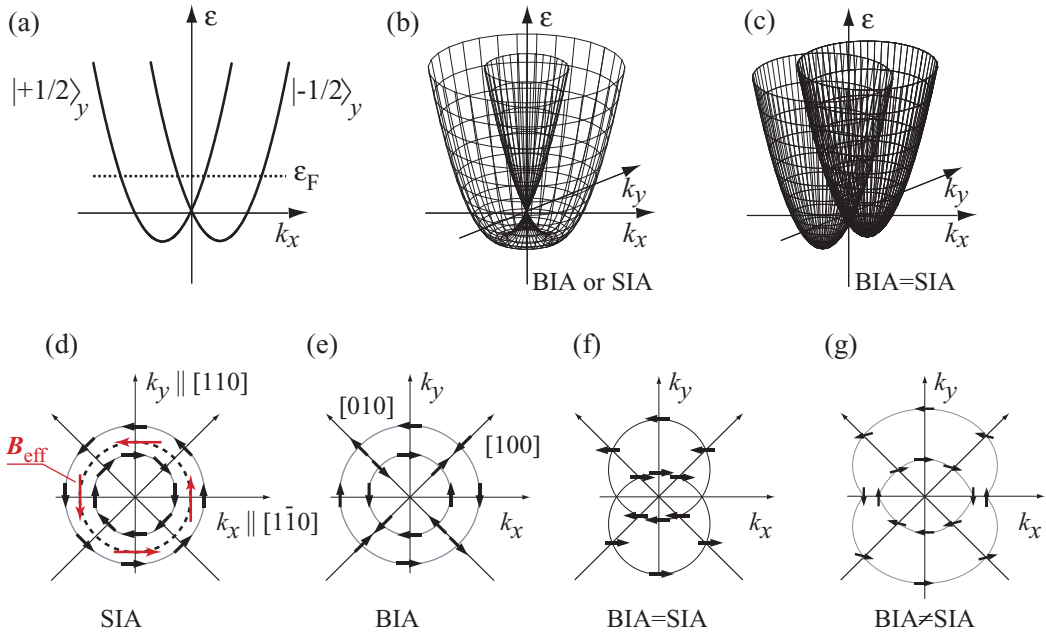}
\end{center}
\caption{
Panel (a) illustrates the SIA/BIA spin splitting due to $k_x \sigma_y$ terms in the 
effective Hamiltonian, here $|\pm 1/2 \rangle_y$ label the eigenstates with fixed $y$ spin components. 
Panels (b) and (c) show schematic 2D  band structure with {\boldmath$k$}-linear
terms for C$_{2v}$ symmetry.
The energy $\varepsilon$ is plotted as a function of $k_x$ and $k_y$ in 
(b) with only one type of inversion asymmetry, BIA or SIA, and in (c)
for equal strength of the BIA and SIA terms in the Hamiltonian.
The bottom panels show the distribution of spin orientations at
the Fermi energy for different strengths of the BIA and SIA
terms. After \protect \cite{review2003spin,PRL04}.}
\label{A1spinfig01}
\end{figure*}

\subsection{Rashba/Dresselhaus terms in (001)-grown zinc-blende structure based 2DES}
\label{001}

Quantum well structures made of III-V semiconductors MBE grown on (001)-oriented 
substrates are the most studied low dimensional systems. 
The point symmetry group of 
these structures can be either D$_{2d}$ or C$_{2v}$ which both belong 
the  gyrotropic point groups~\cite{gyrotropy}
and, consequently allow linear in wavevector spin splitting.
The $D_{2d}$ symmetry corresponds to  (001)-oriented symmetrical  quantum wells. 
In such QWs only BIA terms may exist. 
If an additional up-down asymmetry is present due to, e.g. nonequivalent
interfaces, asymmetric doping or electric field applied normally to QW plane, 
then the symmetry is reduced from D$_{2d}$ to C$_{2v}$ giving rise to SIA. 
For these QWs
the tensor elements can be conveniently presented 
in the coordinate system $(xyz)$ with $x\parallel [1\bar{1}0]$, $y\parallel [110]$,
$z\parallel [001]$, see Table~\ref{table1} and Fig.~\ref{planes001}a. 
The coordinates $x$ and $y$ lie  in the reflection planes $m_1$ and 
$m_2$ of both point groups and are perpendicular to the
principal two-fold rotation axis C$_2$,
see Fig.~\ref{planes001}b showing symmetry elements for QWs of C$_{2v}$ point group.

For $D_{2d}$ point symmetry the linear in $\bm k$ wavevector spin splitting is given by
\begin{eqnarray} 
\label{A1currentequ6}
{H}_{\rm BIA} = \beta (\sigma_xk_y +\sigma_y k_x),
\end{eqnarray}
where $\beta$ is called the (2D) Dresselhaus constant. It follows 
from Eq.~\eqref{Dress_bulk} that the substantial contribution to $\beta$ 
comes from the bulk spin-orbit 
coupling, which, taking into account that for confined electrons $\langle k_z \rangle$ becomes zero but
$\langle k_z^2\rangle$ does't,  yields  $\beta = -\gamma \langle k_z^2\rangle$. Here 
the brackets mean averaging over the size-quantized motion~\cite{A1Dyakonov86p110}.
It is important to note, that historically many authors use coordinate axes directed 
along cubic axes, i.e. 
$x'\parallel$~[100] and  $y'\parallel$~[010]. As in this coordinate system 
$x'$ and $y'$ are tilted by 45$^\circ$
to the mirror planes the other spin and $\bm k$ components are mixed and the form 
of the Hamiltonian changes. 
In this case, we have widely used in the literature form of 
${H}_{\rm BIA} = \beta (\sigma_{x'}k_{x'} - \sigma_{y'} k_{y'})$.

In the asymmetric  QWs, belonging to C$_{{\rm 2v}}$ point group 
and having nonequivalent $z$ and $-z$ directions, SIA gives rise to 
additional terms in $H_{\rm SO}$
so that  now $H_{\rm SO}  = {H}_{\rm BIA}+ {H}_{\rm SIA}$. 
The form of ${H}_{\rm BIA}$ remains unchanged, see Eq.~\eqref{A1currentequ6},
and the SIA term assumes the form
\begin{eqnarray} \label{A1currentequ7}
{H}_{\rm SIA} = \alpha (\sigma_xk_y -\sigma_y k_x) = \alpha(\bm \sigma \times \bm k)_z,
\end{eqnarray}
where $\alpha$ is called the Rashba constant.
Obviously the form of  this term is independent of the
orientation of Cartesian coordinates in the plane of the QW.
Equations~\eqref{A1currentequ6} and~\eqref{A1currentequ7}
show that linear in wavevector band spin splitting is possible for in-plane spin
components only. This fact can be illustrated by simple symmetry arguments. 
It follows from the Neumann's Principle 
that the linear in $\bm k$ spin splitting 
can only occur  
for those components of $\bm{k}$ for which 
there are components of the pseudovector $\bm{S}$ 
(the corresponding quantum-mechanical operator is $\bm \sigma/2$)
transforming in the same way.
Let us illustrate  
it  for spin aligned  along $y$-direction, i.e., for $S_y$. 
Figure~\ref{planes001}(c) shows the  symmetry elements of 
asymmetric QWs (point group C$_{{\rm 2v}}$) together with the transformation of $k_x$ and $S_y$ 
by the mirror reflection in $m_1$ plane. 
We see, that the reflection in the plane $m_1$, as well as in $m_2$,
transforms the wavevector component $k_y$ and the pseudovector component $S_x$ 
in the same way: 
$k_x \rightarrow k_x$, $S_{y} \rightarrow S_{y}$ for the plane $m_1$ (see
figure~\ref{planes001}(c)) and $k_x \rightarrow -k_x$, $S_{y}
\rightarrow - S_{y}$ for the plane $m_2$. 
As the remaining  C$_2$-axis also transforms $k_x$ and $S_y$
equally  the linear in $\bm k$ spin splitting 
connecting these components 
becomes  possible yielding the $k_x \sigma_y$ terms in the 
effective Hamiltonians~\eqref{A1currentequ6} and~\eqref{A1currentequ7}.
The corresponding band structure is sketched in Fig.~\ref{A1spinfig01}(a).
Similar arguments hold for 
$k_y$ and $S_{x}$ ($k_y \sigma_x$ terms), but not for the out-of-plane component $S_{z}$. Consequently,
the linear in $\bm k $ spin splitting for out-of-plane spin 
is forbidden by symmetry.

The distribution of spin orientation in the states with a given $\bm k$
can be visualized by writing the spin-orbit interaction term in
the form 
\begin{equation}
\label{Beff}
H_{\rm SO} = \bm \sigma \cdot \bm B_{eff}(\bm{k}),
\end{equation}
where
$\bm B_{eff}(\bm{k})$ is an effective magnetic field (with absorbed Bohr magneton
and $g^*$-factor~\cite{Sinitsin}), which provides the relevant 
quantization axes.
%
The index ''effective'' indicates that $\bm{B}_{eff}(\bm{k})$ 
is not a real magnetic field because it does not break the
time-inversion~\cite{Omega}. Consequently, in the presence of SIA/BIA spin splitting 
the Kramers-relation $\varepsilon (\mbox{\boldmath$k$}, \uparrow) =
\varepsilon(-\mbox{\boldmath$k$}, \downarrow)$ holds.
By comparison of Eq.~\eqref{Beff} with Eqs.~\eqref{A1currentequ6} and~\eqref{A1currentequ7} 
one obtains for pure SIA ($\beta $=0) and pure BIA ($\alpha $=0) 
the effective magnetic fields in forms
\begin{equation}
\label{Beff1}
    \bm B_{eff}^{\rm SIA} = \alpha (k_y , - k_x ), \qquad
    \bm B_{eff}^{\rm BIA} = \beta (k_y ,k_x ).
\end{equation}
The effective magnetic field and spin orientations 
for, both, Rashba and Dresselhaus coupling are schematically 
shown by arrows in Fig.~\ref{A1spinfig01} (d) and (e), respectively. 
Here it is assumed for concreteness that $\alpha, \beta >0$. 
For the SIA  case the effective magnetic field and,
hence, the electron spin in the eigenstates with the wavevector $\bm k$ 
are always perpendicular to the 
{\boldmath$k$}-vector, see Fig.~\ref{A1spinfig01}(d). 
By contrast, for the BIA contribution, the angle between {\boldmath$k$}-vector
and spins depends on the direction of {\boldmath$k$}, see Fig.~\ref{A1spinfig01}(e). 
In the presence of both  SIA and  BIA spin-orbit couplings (C$_{2v}$ symmetry) 
the $[1\bar {1}0]$ and the [110] axes become strongly non-equivalent. For $\bm k \parallel [1\bar {1}0]$ 
the eigenvalues of the Hamiltonian are then given by ${\varepsilon_\pm =
\hbar^2 k^2 / 2m^*\pm (\alpha - \beta
)k}$ and for $\bm k \parallel
[110]$ by ${\varepsilon_\pm = \hbar^2 k^2 / 2m^*\pm
(\alpha + \beta )k}$. 
For an arbitrary direction of $\bm k$,
the energy spectrum of such systems consists of two branches with
the following anisotropic dispersions~
\begin{equation}\label{spectrum}
    \varepsilon_\pm({\bm k}) = {\hbar^2 k^2 \over 2 m^*} \pm
    k\sqrt{\alpha^2 + \beta^2 + 2 \alpha \beta \sin{2 \vartheta_{\bm k}}} \,\, ,
\end{equation}
where $\vartheta_{\bm k}$ is the angle between $\bm k$ and the $x$ 
axis~\cite{A1Silva92,Sandoval2013}.
The energy dispersion for  {\boldmath$k$}-linear SIA, BIA and combined SIA/BIA terms 
is illustrated in Fig.~\ref{A1spinfig01}(a)--(c).  
In the case of BIA only ($\alpha = 0$) or SIA only
($\beta=0$) the band structure is the result of the
revolution around the energy axis of two parabolas symmetrically
displaced with respect to {\boldmath$k$} = 0. 
The interplay of SIA and BIA is illustrated in  panels (d)-(g). 
If the strengths of BIA and SIA are the same then the 2D band structure consists of
two revolution paraboloids with revolution axes symmetrically
shifted  in opposite directions with respect to {\boldmath$k$} = 0
(Fig.~\ref{A1spinfig01}~(c)). Now all spins are oriented along
$\pm x$-axes as shown in Fig.~\ref{A1spinfig01}~(f). In
Fig.~\ref{A1spinfig01}~(g) we have shown a constant energy surface
and direction of spins for $\alpha \neq \beta$.

So far we discussed only $\bm k$-linear terms in the Hamiltonian. In fact, in
zinc-blende structure based (001)-grown quantum wells also terms cubic in
$\bm k$ are present which stem from the Dresselhaus term in 
the host bulk material, Eq.~\eqref{Dress_bulk}. A cubic contribution  modifies the 
Dresselhaus spin-splitting yielding
%
\begin{equation}
\label{HBIAcub}
H_{\rm BIA}^{cub}  =    
{\gamma\over 2} \left(\sigma_x k_y - \sigma_y k_x \right) (k_y^2-k_x^2). 
\end{equation}
These terms influence some of spin dependent phenomena like spin relaxation or WAL and
should be taken into account in particularly in narrow band materials and highly doped QWs as well as at high temperature. 
The corresponding $\bm k$-cubic effective magnetic field 
defined via $H_{\rm BIA}^{cub} =   \bm{\sigma} \cdot {\bm B}_{\rm cub}$
can be conveniently decomposed into ${\bm B}_{\rm cub}={\bm B}_{\rm cub}^{(1)}+{\bm B}_{\rm cub}^{(3)}$, 
where~\cite{omega2}
\begin{equation}
\label{Beff3}
    {\bm B}_{\rm cub}^{(1)} = -\frac{\gamma k^2 }{4} (k_y ,k_x), \quad 
    {\bm B}_{\rm cub}^{(3)} = \frac{\gamma k^3 }{4} \left( \sin 3\vartheta_{\bm k}, -\cos 3\vartheta_{\bm k}  \right) \,  .
\end{equation}
As the effective magnetic fields  $\bm B_{eff}^{\rm BIA}$ and ${\bm B}_{\rm cub}^{(1)}$ 
containing the first-order Fourier harmonics ($\propto \sin{\vartheta_{\bm k}}$ and $\cos{\vartheta_{\bm k}}$)
have the same form [see Eqs.~\eqref{Beff1} and~\eqref{Beff3}] they can be combined 
as
\begin{equation}
\label{Beff1n}
    \bm B_{eff}^{(1)}=\bm B_{eff}^{\rm BIA}+{\bm B}_{\rm cub}^{(1)} = \tilde{\beta}(k_y ,k_x)
\end{equation}
with the renormalized Dresselhaus constant 
\begin{equation}
\label{betatilde}
\tilde{\beta} = \beta - \gamma k^2/4.
\end{equation}
Note that the term $\gamma k^2/4$ scales with $k$ which 
for equilibrium electron gas is equal to the Fermi wavevector and
respectively, scales with the electron density. In GaAs heterostructures
cubic in $\bm k$ terms are usually unimportant and 
in the further consideration we will use $\tilde{\beta}$
for analysis of narrow band semiconductors only.

\subsection{Rashba/Dresselhaus terms in (110)-grown zinc-blende structure based 2DES}
\label{110}

Quantum wells  on (110)-oriented GaAs substrates attracted 
growing attention  due to their extraordinary slow 
spin dephasing which can reach  several hundreds of 
nanoseconds~\cite{Ohno1999,Harley03,Dohrmann2004,Hall2005,PRL08110,Mueller2008,Mueller10,Griesbeck2012}.
As addressed above, the reason for the long spin lifetime in this type of QWs is their
symmetry: in (110)-grown QWs the BIA effective magnetic 
field $\bm B_{eff}(\bm{k})$ points into the growth 
direction~\cite{A1Dyakonov86p110} 
therefore spins oriented along this direction do not precess. 
Hence the Dyakonov-Perel spin relaxation mechanism, 
which is based on the spin precession in $\bm B_{eff}(\bm{k})$  
and usually limits the spin lifetime of conduction electrons, 
is suppressed.

\begin{figure}[tb]
\begin{center}
\includegraphics[width=\linewidth]{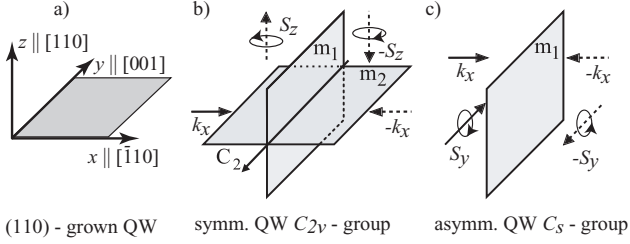}
\end{center}
\caption{(a) Coordinate system of the (110)-grown III-V QW. 
(b) mirror planes $m_1$ and $m_2$ and C$_2$-axis in symmetric QW
belonging to C$_{2\mathrm{v}}$ point group. 
(c) Remaining symmetry elements in asymmetric QWs (C$_s$ point group).
Arrows in the drawing (b) show  that the
reflection in the mirror plane $m_1$ changes the sign of both the
polar vector component $k_x$ and the axial vector component $S_z$, 
demonstrating that a linear coupling $k_x$ and $S_z$ is 
allowed under these symmetry operation. This
coupling is also allowed by the other symmetry 
operations (mirror reflection by the plane $m_2$
and the C$_2$ rotation axis) of the point group yielding the $k_x \sigma_y$ 
terms in the effective Hamiltonian. 
By contrast the  in-plane spin $\bm S$ at reflection in plane $m_2$
transforms differently compared to any  in-plane wavevector components.
Arrows in the sketch (c) demonstrate that in asymmetric QWs for which $m_2$
is removed, the coupling between the in-plane spin $S_y$  and wavevector $k_x$
becomes possible. Same arguments are valid for the coupling of $S_x$ and $k_y$.
Thus SIA in asymmetric QWs results in the in-plane effective magnetic field
and gives rise to the Dyakonov-Perel spin relaxation for 
spins oriented along growth direction.
}
\label{figure01}
\end{figure}

Depending on the equivalence or nonequivalence of the QW interfaces, 
i.e. presence or absence of SIA,  the structure symmetry may belong to one of the point groups:
C$_{2v}$ or C$_s$, respectively. 
While the point group symmetry 
of symmetric (110)- and asymmetric (001)-oriented III-V quantum wells is the same
(C$_{2v}$) the Dresselhaus spin splitting links different 
components of electron spin and wavevector.  The reason for this fact, 
strange on the first glance,  is that by contrast to (001)-oriented QWs for which 
mirror reflection planes $m_1$ and $m_2$ are oriented normal to the QW plane
(see Fig.~\ref{planes001}), in symmetric (110) QWs one of the planes, say $m_2$, 
coincides with the plane of QW. The symmetry elements of symmetrical 
(110)-grown QWs are shown in Fig.~\ref{figure01}b.
By simple symmetry analysis we find that the only wavevector 
and spin components transforming in the same way are  $k_x$ and $S_{z}$, 
i.e. the effective magnetic field caused by the spin splitting 
points along the growth  axis. The reflection of these components 
in the $m_1$ mirror plane resulting in 
$k_x \rightarrow -k_x$, $S_{z} \rightarrow -S_{z}$,  
are shown in Fig.~\ref{figure01}(b). 
The corresponding Hamiltonian has the form
\begin{equation} 
\label{BIA_Cs}
{H}_{\rm BIA} = \beta \sigma_z k_x.
\end{equation}

\begin{figure*}[tb]
\begin{center}
\includegraphics[width=0.7\linewidth]{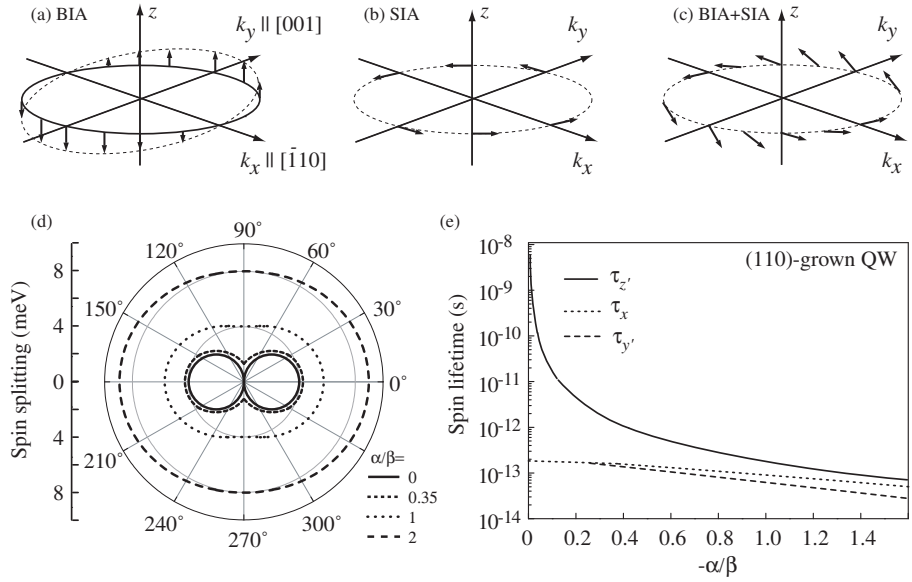}
\end{center}
\caption{(a-c)
Crystallographic directions and effective magnetic fields 
for the lower conduction subband in (110)-grown QW. 
Plots (a)-(d) sketches $B_{eff}(\bm{k})$  for BIA, SIA and SIA/BIA 
interplay, respectively.
Note that for $\alpha$=0, panel (a), $B_{eff}(\bm{k})$ is aligned along $z$ axis
for any $\bm k$ and states with wavevector $k_y$ are spin degenerated. 
(d) Calculated  values of the spin splitting for states lying on a circle 
in the $k_x$---$k_y$ plane with $k=0.01 \AA^{-1}$ for 
various $\alpha$/$\beta$ ratio. Here $\beta$ is {10$^{-9}$~eV cm} is assumed for calculations.
(e) Spin lifetimes as a function of the ratio of the SIA
and the BIA parameters. 
The labels $\tau_{x,y',z'}$ refer 
to the spin relaxation tensor in the axes $x$, $y'$, and $z'$.
The latter two lie in ($y,z$) plane and are tilted to the 
axes $y$ and $z$, see Table~\ref{table1},
by the angle $\theta=\arctan{(\alpha/\beta)}$. Data are given after~\cite{Cartoixa2005_2}.
}
\label{cartoix_110_splitting}
\end{figure*}

Structure inversion asymmetry removes the mirror reflection plane $m_2$ 
and enables spin splitting for the in-plane spin components. 
This is illustrated in Fig.~\ref{figure01}(c) 
showing that in asymmetric (110)-grown structures 
$k_x$ and $S_y$  transform equally. 
Additional  terms in the Hamiltonian  caused by
the symmetry reduction have the same form as  the 
Rashba terms in (001)-oriented QWs Eq.~\eqref{A1currentequ7}.
The in-plane effective magnetic field due to Rashba spin-orbit coupling 
results in spin dephasing even for spins oriented along growth direction
and the benefit of (110) QWs disappears~\cite{combined}.
The energy spectrum of such systems consists of two branches with
the following anisotropic dispersions
%
\begin{equation}
\label{spectrum110}
    \varepsilon_\pm({\bm k}) = {\hbar^2 k^2 \over 2 m^*} \pm
    k \sqrt{\alpha^2 + \beta^2 \cos^2{\vartheta}_{\bm k}} \,\,.
\end{equation}
Like in (001) QWs the different forms of the BIA and SIA terms result in their interference
substantially affecting the band spin-splitting.  
The spin splitting and calculated spin relaxation times 
for some  $\alpha$/$\beta$-ratios are shown in Fig.~\ref{cartoix_110_splitting}(d) and (e), respectively.
Moreover, the $\bm k$-cubic terms for symmetric (110) QWs result in the effective magnetic field also 
pointing in the growth direction, 
so they do not lead to spin relaxation of the normal spin component as well.

\subsection{Rashba/Dresselhaus terms in (111)-grown zinc-blende structure based 2DES}
\label{111}

\begin{figure}[tb]
\begin{center}
\includegraphics[width=\linewidth]{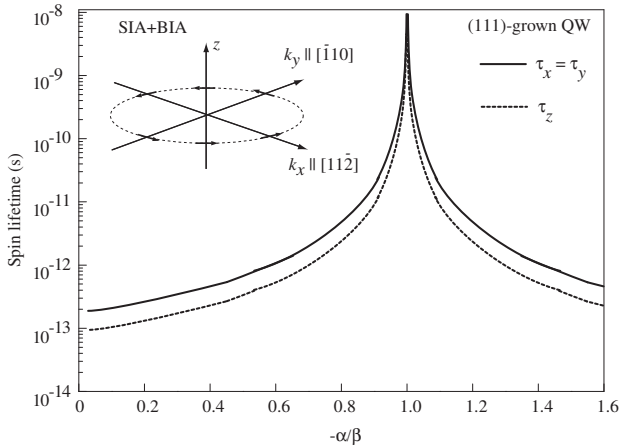}
\end{center}
\caption{ Spin lifetimes in (111)-grown QW as a function of the SIA/BIA-ratio. 
Inset sketches ${\bm B}_{eff}(\bm{k})$  for QWs with any values of SIA/BIA ratio.
Note that in the  case of $\beta  =- \alpha $ effective magnetic field ${\bm B}_{eff}(\bm{k})=0$ for any electron wavevector
and states become spin degenerated. 
Data are given after~\cite{Cartoixa2005_2}.
}
\label{cartoix_111_splitting}
\end{figure}

Quantum wells grown along [111]-direction draw attention primary 
due the possible suppression of the Dyakonov-Perel spin relaxation mechanism 
for all spin components~\cite{Balocchi2011,Ye2012,Biermann2012,Hernandez2012,Wang2013,Cartoixa2005_2,Vurgaftman2005,Sun2010}. 
The reason for this interesting feature is the 
formal identity of the $\bm k$-linear  Dresselhaus and Rashba Hamiltonians, 
which both have a form of  Eq.~\eqref{A1currentequ7} but imply different 
constants, $\beta$ and $\alpha$, respectively~\cite{A1Dyakonov86p110,Cartoixa2005_2}.
As a result, the total spin-orbit Hamiltonian 
can be written in form
\begin{eqnarray} \label{A1currentequ7_111}
{H_{\rm SO}} = (\beta + \alpha)(\sigma_x k_y -\sigma_y k_x).
\end{eqnarray}
The corresponding effective magnetic fields are shown in the inset in Fig.~\ref{cartoix_111_splitting}.
A straightforward consequence of Eq.~\eqref{A1currentequ7_111} is
that conduction band becomes spin degenerate to first order in $\bm k$ 
for \textit{any}  wavevector direction in the case that 
BIA and SIA coefficients would have equal magnitude 
but opposite signs, i.e. ($\beta=-\alpha$). 
The most significant feature of this configuration is that the spin
lifetimes would become tremendously increased.  The dependence of the 
in-plane and out-of-plane spin relaxation times on the ratio between the 
Rashba and Dresselhaus coefficients is shown in Fig.~\ref{cartoix_111_splitting}. 

\subsection{Rashba/Dresselhaus terms in (113)-, (112)-,  (013)-, and \textit{miscut} (001)-grown 
zinc-blende structure based 2DES}
\label{113}

So far we discussed widely spread configurations of zinc-blende structure 
based quantum wells. To be complete we also address the band spin splitting 
in QWs grown in more exotic directions. 
These are (113)-, {(112)-,}  
and (013)-oriented 2DES as well as  (001) \textit{miscut} structures. 
The former orientation is usually used for growth of high mobility $p$-type GaAs QWs, 
which attracted notable attention due to the possibility to obtain long spin 
relaxation times~\cite{Ferreira91p9687,Bastard93p439,PRL2002bleching,Petra2004,Korn2010,Korn2010_2,Kugler2009}.
Quantum wells of (112)-  and (013)-orientations are mainly used for growth of HgTe 2DES, 
which become particularly important since discovery of the topological insulator states
in this material~\cite{Koenig2007,Zhang}
openning the possibility to study  physics of Dirac fermions in the systems 
with the strong spin-orbit interaction. 
Miscut heterostructures are usually MOCVD-grown 
on slightly tilted (001) substrates.

Zinc-blende structure based  
(113)-, (112)- and \textit{miscut} (001)-oriented 2DES belong to the 
symmetry point group
C$_s$, which contains only two elements, the identity and one
mirror reflection plane, $m$, being normal to the 2DES plane. 
A natural coordinate system for these structures grown in $z$-direction
are $x$ - normal to the plane $m$ and 
$y$ - orthogonal to $x$ and $z$, cf. Table~\ref{table1} for (113)-grown QWs. 
Then BIA spin-splitting is given by
\begin{eqnarray} \label{BIA_cs}
{H}_{\rm BIA}^{C_s} = \beta_1 \sigma_xk_y + \beta_2\sigma_y k_x + \beta_3 \sigma_{z}k_x \, \, ,
\end{eqnarray}
while the SIA Hamiltonian is again described by a universal form of Eq.~\eqref{A1currentequ7}.
A particular feature of such 2DES compared to (001)-grown QWs is the
appearance of the spin splitting for spins oriented along the growth
direction.
Note that the spin-orbit coupling terms in asymmetric (110) QWs, which 
have C$_s$ symmetry, are also described by 
the sum of the Hamiltonians Eqs.~\eqref{A1currentequ7} and~\eqref{BIA_cs}.

Quantum wells prepared on (013)-oriented substrates
belong to the trivial point group C$_1$
lacking any symmetry operation except the identity. 
This is true even for structure-symmetric QWs, 
where SIA is absent.
Hence, symmetry does not impose any restriction on the 
relation between the spin and
wavevector components:
\begin{equation}
    H_{\rm SO}^{C_1} = \sum_{lm} \Lambda_{lm} \sigma_l k_m.
\end{equation}
Here $l=x,y,z$ and $m=x,y$. 
Since all components of the 
pseudotensor $\bm{\Lambda}$ may be different from zero the 
spin splitting is allowed for any relative directions of 
spin and electron wavevector.
Moreover, the ratio between the components of the tensor 
$\bm{\Lambda}$ can be changed in nontrivial way (including the sign inversion)
by varying the experimental conditions like e.g. sample temperature 
or carrier density.

To complete the picture of spin splitting in zinc-blende structure based QWs  
we note that symmetry reduction can also be obtained by e.g.
applying stress,
fabricating ungated/gated 
lateral superlattices,
and growing quasi one-dimensional wires.
The form of the spin-orbit Hamiltonian in such  structures depends on the resulting symmetry 
(C$_{2v}$, C$_s$ or C$_1$)
and is described by the corresponding equations discussed above.

\subsection{Wurtzite-type semiconductor structures}
\label{sectionwurtzite}

Wurtzite-type bulk semiconductors, like GaN or InN,
are described by a non-symmorphic space group 
C$_{6v}^{4}$ containing a non-trivial 
translation. However,  the physical effects are determined
 by the point-group symmetry. 
The point group of wurtzite semiconductors  C$_{6v}$
is gyrotropic~\cite{gyrotropy} and, therefore, allows the linear 
in wavevector spin splitting. As it was pointed out in 
Ref.~\cite{A1Rashba60p1109} in these media the spin-orbit part of the Hamiltonian has the form
\begin{eqnarray} \label{GaN}
{H}_{\rm bulk} =  \beta (\bm \sigma \times \bm k)_z,
\end{eqnarray}
where the constant $\beta$  is solely due to BIA. 
Here $z$-axis is directed along the hexagonal $c$-axis~\cite{cubicGaN}.
In heterostructures, an additional
source of $\bm k$-linear spin splitting, induced by SIA becomes possible. 
If both, bulk and structure asymmetries, are present the resulting coupling
constant is equal to the sum of BIA and SIA contributions
to the spin-orbit part of the Hamiltonian. Thus, for 
2DES based, e.g., on GaN or InN grown in (0001) direction, 
the total  spin-orbit part of the Hamiltonian has exactly the same 
form as that for (111)-grown zinc-blende based structures 
discussed in section~\ref{111}:
\begin{eqnarray} 
\label{wurtzit}
H_{\rm SO} = (\beta + \alpha)(\sigma_x k_y -\sigma_y k_x) \, ,
\end{eqnarray}
and for $\beta  =- \alpha $ conduction bands become spin degenerate 
to first order in $\bm k$ for \textit{any}  wavevector direction.

\subsection{SiGe QWs} 
\label{SiGe}

Finally, we briefly discuss SiGe QWs. 
Since both Si and Ge possess inversion center 
SiGe heterostructures do not have BIA. 
However, both IIA, with a BIA-like form of the Hamiltonian~\cite{A1Roessler02p313}, and SIA may lead to {\boldmath$k$}-linear terms~\cite{A1PRB02SiGe,A1Wilamowski02p195315,A1Golub03bialike,Nestoklon2006,Nestoklon2008,Prada2011}.
The symmetry of Si/(Si$_{1-x}$Ge$_x$)$_n$/Si QW in the absence of SIA
depends on the number~$n$ of the mono-atomic layers  in the well. In the case of
(001)-grown  QW structures with an
even number~$n$, the symmetry of  QWs is $D_{2h}$ which is
inversion symmetric and does not yield {\boldmath$k$}-linear
terms. An odd number of~$n$, however, interchanges the
$[1\bar{1}0]$ and $[110]$ axes of the adjacent barriers and
reduces the symmetry to D$_{2d}$~\cite{A1PRB02SiGe,A1Golub03bialike} 
with the same implication treated above for zinc-blende structure based QWs,
see Eq.~\eqref{A1currentequ6} for IIA.
The symmetry reduction of SiGe structures  to C$_{2v}$
may  be caused by e.g. an electric field (external or built-in) applied along the growth direction.
If the structure is grown along the low-symmetry axis $z
\parallel [hhl]$ with $[hhl] \neq$ [001] or [111], the point group
becomes C$_s$ (see {\it e.g.}~\cite{A1PRB02SiGe}) and contains only two
elements, the identity and one mirror reflection plane
$(1\bar{1}0)$. Here the spin splitting is described by equations presented
in section~\ref{113}.

\section{Determination of SIA and BIA spin splittings by photogalvanic measurements}
\label{method}

A direct way to explore the BIA and SIA interplay, which does not require 
knowledge of microscopic details, is based on the phenomenological 
equivalence of Rashba/Dresselhaus linear in $\bm k$
spin splitting with other phenomena also described 
by a linear coupling of a polar vector, like current, and 
an axial vector, like electron spin.
Indeed, all these effects are described by  second rank
pseudo-tensors whose irreducible components differ by a scalar
factor only and, consequently, are characterized by the same anisotropy.
We discuss here three experimentally accessible effects 
which belong to a large class of 
photogalvanic phenomena~\cite{IvchenkoGanichev,A1book,A1book2,Belinicher,sturman}
and are also  described by such second rank pseudo-tensors. 
In considered here 2DES these effects  are: 
the spin-galvanic effect (SGE)~\cite{Ivchenko89p175,Nature02} 
given by
\begin{equation}
    j_l =  \sum_{m}Q_{l m} S_m \,\, ,
    \label{SGE0}
\end{equation}
the circular photogalvanic effect (CPGE)~\cite{PRL01,Ganichev02} with
\begin{equation}
    j_l= \sum_{m}\chi_{l m}  \hat{e}_m P_{\rm circ} \,\, ,
\end{equation}
and magneto-gyrotropic effect (MPGE) excited by normal incident 
unpolarized radiation which is 
described by~\cite{LechnerAPL2009,BelkovJPCM}
\begin{equation}
\label{unpol}
j_l=  \sum_{m}\xi_{l m} B_m |\bm E|^2 \,\, .
\end{equation}
Here $\bm S$ is the average spin, $P_{\rm circ}$ is a helicity of radiation,
$\hat{\bm e}$ is the unit vector pointing in
the direction of light propagation, 
{$\bm E$} is the complex amplitude of the electric field
of the electromagnetic wave, and {$\bm B$} is an external magnetic field. 
In analogy to the band spin-splitting given by
\begin{equation}
H_{\rm SO}=\sum_{lm}\Lambda_{lm}\sigma_lk_m, 
\end{equation}
where $\Lambda_{lm}$ is a second rank pseudo-tensor,  
these currents can be decomposed into BIA and SIA 
contributions which can be measured separately.  
Therefore, the equivalence of the invariant irreducible 
components of the pseudo-tensors ${\bm \Lambda}$, 
${\bm Q}$, ${\bm  \chi}$  and ${\bm  \xi}$ can be used to 
evaluate the ratio between 
SIA and BIA strength as well as
to determine their relative 
sign~\cite{PRL08110,PRL04,PRB07,LechnerAPL2009,Zhao07GaN,Golub2005,Kohda2012}.
Note that the same arguments are valid for the inversed SGE~\cite{inversedSGE}.

This can be illustrated on example of the spin-galvanic 
effect  where an electric current  is caused by asymmetric  
spin relaxation of  non-equilibrium  spin polarized carriers 
in the system with a spin-orbit splitting of the 
energy spectrum, for reviews see~\cite{review2003spin,GanichevPrettl,A1book2}.
While in general the spin-galvanic effect  does not need optical excitation 
the SIA/BIA interplay can most convenient be studied applying circularly 
polarized radiation for spin orientation of carriers resulting in $\bm S$. 

\begin{figure}
\includegraphics*[width=\linewidth]{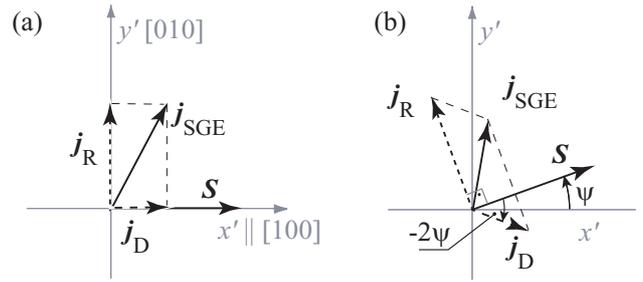}
\caption{ Spin-galvanic currents and its SIA and BIA components 
in a (001)-grown  QW. (a) for the in-plane average spin direction aligned along $x'$,
(b) for $\bm S$ given by arbitrary angle $\Psi$. After~\cite{PRL04,PRB07}.
} 
\label{fig2}
\end{figure}

The SGE current ${\bm j}^{SGE}$ is linked to 
the average spin by a second rank pseudo-tensor $\bm Q$, see Eq.~\eqref{SGE0}, 
and can be presented via the parameters of spin-orbit splitting in (001) QWs  as follows
\begin{equation}
\label{SGEBIASIA}
    j^{\rm SGE}_x = Q (\beta - \alpha)S_y,
    \qquad
    j^{\rm SGE}_y = Q (\beta + \alpha)S_x,
\end{equation}
where $Q$ is a constant determined by the
kinetics of the SGE, namely by the characteristics 
of momentum and spin relaxation processes.
In the coordinate system with cubic axes
($x' \parallel [100]$, $y' \parallel [010]$) 
Eq.~\eqref{SGE0} can be conveniently presented in the form
\begin{equation}
\label{SGE} 
{\bm j}^{SGE}  = Q \left( {{\begin{array}{*{20}c}
 \beta \hfill & -\alpha \hfill \\
  \alpha  \hfill & - \beta \hfill \\
\end{array} }} \right){\bm S} \,\, ,
\end{equation}
demonstrating that for spin aligned along $x'$ or $y'$ 
measurements of the SGE current parallel and perpendicular to $\bm S$ 
directly yield the $\alpha/\beta$-ratio.
This is sketched in Fig.~\ref{fig2}(a) showing the average spin ${\bm
S} \parallel x'$ and the spin-galvanic current ${\bm j}^{SGE}$, which is
decomposed into $ j_{x'}= j_D$ and $j_{y'}= j_R$ proportional to the
Dresselhaus  constant $\beta$ and the Rashba constant $\alpha$, respectively.
This configuration represents the most convenient experimental geometry
in which the ratio of the currents measured 
along  $x'$- and $y'$-axes yields
\begin{equation}
\label{RDgeomI}
\frac{\alpha}{\beta} = \frac{j_{y'}(\bm S \parallel {x'})}{j_{x'}(\bm S \parallel {x'})}\,\,.
\end{equation}
We emphasize that this geometry unambiguously shows whether 
the Rashba or  Dresselhaus contribution is dominating. 
Furthermore, these measurements provide experimental determination 
of both the ratio and the relative sign of 
the Rashba and Dresselhaus constants.

Analogously to the spin splitting, symmetry arguments 
yield that for arbitrary orientation of the average spin ${\bm S}$ the current 
${\bm j}_R$ is always perpendicular to ${\bm S}$ 
while the current ${\bm j}_D$ encloses an angle $-2\Psi$ with ${\bm S}$, where
$\Psi$ is the angle between ${\bm S}$ and the $x'$-axis.  
The strength of the total current $j^{SGE}$ is given by 
the expression
\begin{equation}
\label{j_total}
j^{SGE} = \sqrt{j_R^2 + j_D^2 - 2 j_R j_D \sin{2\Psi}} \,\, ,
\end{equation}
which has the same algebraic form as the spin-orbit term in the
band structure, see Eq.~\eqref{spectrum}. 
Taking the ratio between Rashba and Dresselhaus current contributions 
cancels 
the scalar factor $Q$, which contains all microscopic details~\cite{PRL04,PRB07}.
Hence, by mapping the magnitude of the photocurrent in the plane of the QW the 
$\alpha/\beta$-ratio can be directly extracted from experiments. 
Similar consideration can be made for the circular photogalvanic
and magneto-gyrotropic effects. The details of 
the method can be found 
in Ref.~\cite{PRL08110,PRL04,PRB07,LechnerAPL2009,IvchenkoGanichev,Kohda2012,belkovganichevreview,BelkovGanichev}.

An important advantage of the discussed method
is that it applies the photogalvanic effects, 
which are very general and have been detected in a large variety of low dimensional semiconductor 
structures of a very different designs, for resent reviews on 
these effects see~\cite{IvchenkoGanichev,A1book2,GanichevSchliemann,BelkovGanichev,Winkler06p0605390},  
including topological insulators and other systems with Dirac 
fermions~\cite{Karch2010,Romashko2010,Karch2011,Hosur2011,Kvon2011,McIverTI,Dora_PGE_TI,Wu2012,Drexler2013,PRB_HgTe,Artemenko2013}.
The studies of the last decade show that these
effects in 2DES can be detected in a wide temperature range 
including technologically important room temperature
and applying radiation in a wide frequency range, from microwaves up to visible light. 
It is important to note that all photogalvanic effects 
addressed above are caused by  the terms $\bm B_{eff}^{(1)}(\bm k)$ and $\bm B_{eff}^{\rm SIA}(\bm k)$
in the effective magnetic field,  which are first angular harmonics 
of $\vartheta_{\bm k}$, see section~\ref{001}.
The rest cubic terms in the effective magnetic field 
$\bm B_{eff}^{(3)}(\bm k) \propto \sin{3\vartheta_{\bm k}}, \cos{3\vartheta_{\bm k}}$ 
do not result in the discussed photogalvanic currents, however, they 
modify the spin-splitting and may
affect spin relaxation and the anisotropy of spin-flip Raman
scattering~\cite{Jusserand95p4707,Miller03,Studer10,A1book,Lusakowski2003,Kettemann2007,Flatte}.
Consequently, photogalvanics based methods 
provide the information on the SIA/BIA terms 
given by Rashba constant $\alpha$ and renormalized 
Dresselhaus constant $\tilde{\beta}$. 

\section{Experimental technique}
\label{technique}

Photogalvanics have been used to probe SIA/BIA interplay in a large variety
of low dimensional structures of different design yielding information
on the modification  of the  SIA/BIA-ratio upon changing of various macroscopic parameters 
like crystallographic orientation, doping position, quantum well widths, temperature etc. 
Zinc-blende and wurtzite semiconductor based  heterostructures as well as 
SiGe quantum wells were studied. For optical excitation a great variety of
radiation sources have been used including pulsed and cw molecular THz 
lasers~\cite{GanichevPrettl,Ganichev95DX,Ziemann2000,GanichevPrettl2002,Kvon2008,Karch2010}, 
free electron lasers~\cite{GanichevPrettl,Knippels99p1578,Svelto2010,PRB2008,Wittmann2008,Wittmann2010,Danilov2009}, 
CO$_2$ lasers~\cite{A1PRB02SiGe,Wittmann2010,Jiang2011_1}, 
Ti-sapphire and other solid state lasers~\cite{Yang06p186605,Frazier2009KhodropPGE,Belkov2003}, 
seiconductor lasers~\cite{Cho2007_1,Dai2010}, 
He-Cd laser~\cite{Zhang2010ZnO},
time-domain THz systems~\cite{GanichevPrettl,Bieler05,Priyadarshi2013,Sakai,Lee}, 
conventional Gunn diodes~\cite{Drexler2010} etc.. 
While SIA/BIA interplay has been studied in a wide frequency range from microwaves 
to the near infrared, microwaves/terahertz radiation are particularly 
suitable for the methods addressed in the previous section.
First of all, in the microwave/terahertz range photogalvanics
may be observed and investigated much more easily than in the 
visible or near infrared ranges, where strong spurious photocurrents, 
caused by other mechanisms like the Dember effect~\cite{GanichevPrettl}, photovoltaic effects at contacts etc., 
mask the relatively weak spin photogalvanic currents. Secondly, in contrast to
conventional methods of optical spin orientation using interband
transitions, terahertz radiation  excites only one
type of charge carrier yielding monopolar spin
orientation, giving the information about spin splitting in one subband.
Furthermore, electrons excited by terahertz radiation remain close
to the Fermi energy which corresponds to the conditions of electric spin injection. 

Obviously photocurrent measurements applying radiation with photon energies smaller than the band gap
require free carriers.
Therefore,  photogalvanic methods are applied to study either doped low dimensional systems
or undoped structures additionally exposed to light resulting in 
the photogeneration of electron-hole pairs. In the latter case photogalvanics is caused
by the superposition of electron and hole contributions which complicates 
the analysis of the spin splitting in a particular band.
Experimental geometry depends on the type of phenomenon used for the SIA/BIA mapping 
(CPGE, SGE or MPGE)  and crystallographic orientation of the studied 
low dimensional structure.  CPGE- and SGE-based methods require 
circularly polarized radiation at oblique, or, for low symmetric structures, normal incidence. 
For MPGE-~\cite{LechnerAPL2009,zerobias} and, in some cases, SGE-based methods~\cite{PRL04}, 
a small external magnetic field is needed. Details of the experimental configurations 
can be found in~\cite{PRL08110,PRL04,PRB07,LechnerAPL2009,IvchenkoGanichev,Kohda2012,belkovganichevreview}. 

In the most frequently used geometry square shaped $3 \div 5$ mm$^2$ size samples 
with  an edge oriented parallel to one of the reflection planes have been studied.
The latter  can in most cases be naturally obtained by cleaving the sample. Several 
pairs of contacts, being needed for electrical measurements, are made in the middle of 
the edges and corners of the squared sample. Although this  geometry of contacts
is sufficient for study of SIA/BIA anisotropy the results accuracy can
be increased by using  a larger number of contact pads forming a circle~\cite{PRL04,Kohda2012}.
The photocurrent ${\bm J}(\theta)$, where $\theta$ is the polar angle, is
measured  in unbiased structures via the voltage drop across 
a 50~$\Omega$ load resistor with a fast storage
oscilloscope or applying standard lock-in technique~\cite{review2003spin}.
We note that a pure optical method to measure photogalvanic currents, which 
provides a unique access to characterization of SIA/BIA in a  contactless way,
has been developed~\cite{Bieler05,Priyadarshi2013,sun2010,Sun:2012ys}. 
It is based on the terahertz emission resulting from
the photogalvanic currents generated by picosecond pulses of near infrared radiation.  
The physical principle is just the same as of the Auston switch\cite{Auston1,Auston2} 
used  for generation of THz radiation  in the terahertz 
time-domain spectroscopy~\cite{GanichevPrettl,Sakai,Lee}.

\section{Interplay of BIA and SIA in (001)-, (110)- and (111)-grown III-V 2D systems}
\label{IIIV}

\subsection{Tuning of structure inversion asymmetry by 
the $\delta$-doping position.
}
\label{001exp}

\begin{figure}
\includegraphics[width=0.5\textwidth]{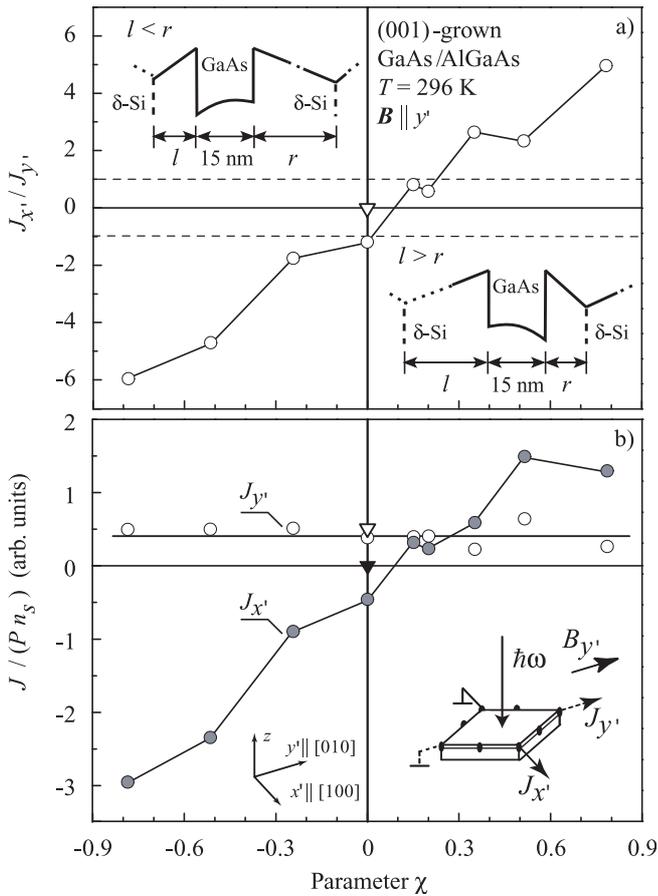}
\caption{ (a) The ratio of the SIA and BIA contributions to 
the MPGE, $J_{x'}/J_{y'}$, as a function of $\chi$.
The triangles show the result for sample grown at 
$T_\delta = 490^\circ$C, the circles demonstrate the 
data for all other samples grown at $T_\delta ~\approx 630^\circ$C. 
Insets show the QW profile and the  doping positions for $l < r$
and for $l>r$.
(b) Dependence of $J/P n_s$ on the parameter $\chi$, here $n_s$ is the carrier density. 
The photocurrents $J_{y'}$ and  $J_{x'}$ are measured along and normal to $\bm B \parallel y'$.
Full and open symbols show $J_x$ and $J_y$, respectively (triangles are the data for sample fabricated with 
reduced temperature during the $\delta$-doping  ($T_\delta = 490^\circ$C)).
Inset shows experimental geometry. After~\cite{LechnerAPL2009}.
}
\label{fig2apl}
\end{figure}

In this section we discuss  the influence of the $\delta$-doping position, 
quantum well width and growth conditions on SIA and BIA in III-V semiconductors 
based (001)-oriented quantum well structures. 
We begin with MPGE investigations of Si-$\delta$-doped 
$n$-type  GaAs$/$Al$_{0.3}$Ga$_{0.7}$As  structures grown by molecular-beam 
epitaxy at typical temperatures in excess of 600$^\circ$C.
The insets in Fig.~\ref{fig2apl} sketch the conduction band  edges of different 
QW structures together with the corresponding $\delta$-doping position.
All QWs have the same width of 15~nm but differ essentially  in their doping profile.
The degree of the doping asymmetry can be conveniently described by
the parameter 
$$\chi = (l-r) / (l+r),$$ 
where $l$ and $r$ are the  
spacer layer thicknesses between QW and $\delta$-layers.
Variation of individual BIA and SIA contributions as well as their 
ratio have been studied applying  magneto-gyrotropic photogalvanic effect~\cite{LechnerAPL2009}. 
In these experiments unpolarized terahertz radiation at normal incidence was used for excitation of
QW structure subjected to an in-plane magnetic field applied along a cubic axis $y'$.
BIA and SIA  photocurrent contributions have been obtained by measuring  the current 
along and perpendicular to the magnetic field, i.e.  $J_{y'}$ and $J_{x'}$, respectively. 
The ratio of SIA/BIA contribution as a function of the parameter $\chi$  is shown in Fig.~\ref{fig2apl}(a) 
demonstrating that it has a strong dependence on the doping position and, moreover, changes 
its sign for $\chi \approx 0.1$. The analysis of the individual contributions 
shown in Figure~\ref{fig2apl}(b) indicates that in all structures  BIA 
remains almost unchanged and the SIA is solely responsible for the 
variation of the band spin splitting with the parameter $\chi$. 

The variation of the parameter $\chi$ shows that SIA   is 
very sensitive to the impurity potential and 
its magnitude and the sign can be controlled by the $\delta$-doping position.
The fact that in nominally symmetric quantum wells with  $\chi = 0$ 
SIA yields a substantial photocurrent signal reflects the  dopant 
migration along the growth direction  (segregation) 
during molecular beam epitaxial growth. This conclusion is supported by the 
MPGE measurements in symmetrically doped sample ($\chi = 0$) fabricated with 
reduced temperature during the $\delta$-doping  ($T_\delta = 490^\circ$C). 
At this conditions segregation is suppressed and  SIA vanishes, see Fig. ~\ref{fig2apl}. 

\begin{figure}
\includegraphics[width=0.5\textwidth]{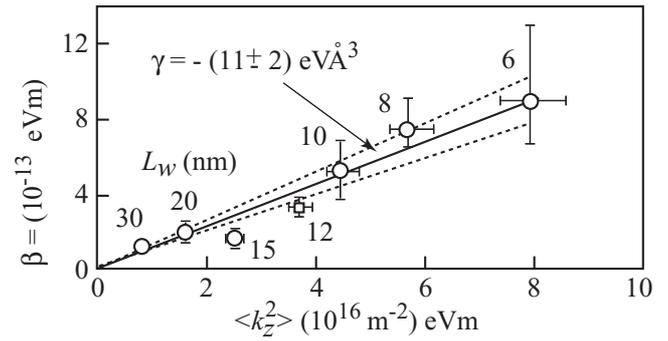}
\caption{ 
Measured linear in $\bm k$ BIA spin splitting, $\beta$, 
vs $\left\langle k_z^2\right\rangle$. Circles and square are the data of Ref.~\cite{Walzer2012} and~\cite{Walzer2012_2}, respectively. 
Solid line is the fit to $-\gamma \left\langle k_z^2\right\rangle$ and 
dotted lines are 95\% confidence interval. Error bars show the estimated uncertainty in the fitted slope. 
Horizontal bars depict $\pm$ 0.5 nm variation in $L_w$ and vertical bars indicate 30\% variation in 
carrier density. After~\cite{Walzer2012}.
}
\label{fig2walser}
\end{figure}

Investigation of the structures with different band profile and $\delta$-doping positions
show  that the largest value of the SIA/BIA ratio is obtained in a single heterojunction~\cite{PRB07}. 
While optical experiments on spin relaxation in undoped samples 
demonstrate that the variation of the band profile does not substantially 
affect SIA~\cite{Eldridge2010_2,Eldridge2011} in doped structures it seems to play an important role.
Indeed in structures with strongly asymmetric potential profile 
like triangular confinement potential or stepped QWs the electron function 
is shifted to one of the interfaces and is strongly affected  
by the impurity Coulomb potential~\cite{shapeQW}. 
The second reason for the enlarged SIA/BIA ratio in wide 2D structures
is the decrease of the Dresselhaus SOI which is given by the size quantization 
of the electron wave vector $k_z$ along the growth direction $z$. 
Theory shows that BIA for a QW of width $L_w$ 
should change after $\left\langle k_z^2\right\rangle \propto 1/L_w^2$~\cite{A1Dyakonov86p110}. 
This behaviour was experimentally confirmed by optical monitoring of the 
angular dependence  of the electron spin precession on the 
direction of electron motion with respect to the crystallographic axes~\cite{Walzer2012,Walzer2012_2}. 
The latter has been obtained driving a current through the structure. 
A set of (001)-grown GaAs/AlGaAs QWs with different well widths 
between 6 and 30 nm and fixed parameter $\chi$ have 
been studied demonstrating a linear increase of the Dresselhaus splitting 
with the increase of the confinement parameter $\left\langle k_z^2\right\rangle$. 
The linear fit presented in Fig.~\ref{fig2walser} yields the 
bulk Dresselhaus coefficient in GaAs,  $\gamma = (-11 \pm 2)$~eV\AA$^3$.
The data also allowed to measure the cubic in $\bm k$
Dreeselhaus term showing that in GaAs it is substantially 
smaller  than the linear one (from 2 up to 30 times,  
for 3~nm and 6~nm QWs, respectively).

\begin{figure}[t]
\includegraphics[width=0.5\textwidth]{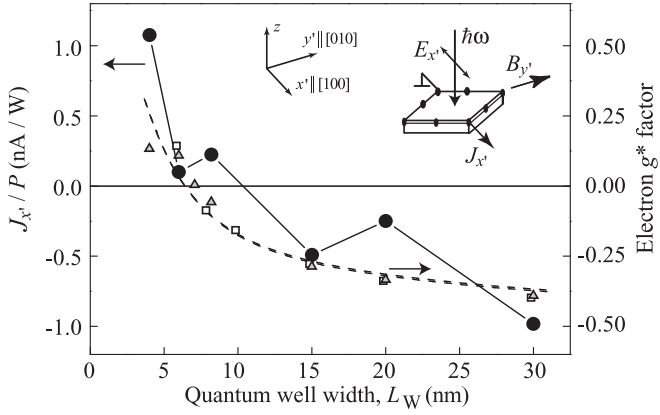}
\caption{Dependence of the MPGE (circles) on $L_{w}$ obtained at 
room temperature, $B_y = \pm 1$~T and photon energy $\hbar \omega = 4.4$~meV 
and corresponding $g^*$-factors (triangles determined by by TRKR~\cite{Lechner2011} 
and squares by comparison of BIA and Zeeman spin splitting~\cite{Walzer2012}).
The inset shows the experimental geometry. After~~\cite{Lechner2011}}
\label{fig2orbit}
\end{figure}

The experiments on the optical monitoring of the 
spin precession as a function of $L_w$
also provided information on the sign of the $g^*$-factor 
confirming its sign change at $L_w =7$~nm~\cite{Walzer2012}.
This inversion is mostly caused by the 
opposite signs of the $g^*$-factor in the GaAs with respect
to the AlGaAs barrier and the fact that for narrow QWs the 
electron wave function deeply penetrates 
into the barrier~\cite{Snelling91,Ivchenko_g_factor,SalisNat01,yugova,Kuglerprb09}. 
The change of sign is of particular importance for the studies of the
magneto-photogalvanic effects resulting from the spin-related roots. As discussed above
the MPGE photocurrent is proportional to the Zeeman 
band spin splitting and is determined by the effective Land\'e factor $g^*$.
The same set of samples as that investigated in the work of Walser et al.~\cite{Walzer2012}
was previously used to provide an experimental evidence for 
spin-related roots of the current formation in most of 
(001) GaAs QWs at room temperature~\cite{Lechner2011}. 
Figure~\ref{fig2orbit} shows  the MPGE photocurrent $\bm{J}^{\rm L}$ as a
function of $L_w$. For comparison, $g^*$ extracted from the  
time-resolved Kerr rotation is also plotted.
As an important result Fig.~\ref{fig2orbit} demonstrates that the photocurrent,
similarly to the $g^*$-factor, changes its sign upon the variation of $L_w$.
However, there is a difference in the zero points: While the $g^*=0$  at $L_{w} = 7$~nm, 
the current vanishes for $L_w \approx  10$~nm.
A small current detected at $g^*$ inversion point, at which 
spin mechanism of MPGE is disabled, is caused by 
the orbital mechanism~\cite{Lechner2011,Tarasenko_orbital},  
which is almost independent on $L_w$. For other 
QW widths the spin related mechanism dominates the total current.
The dominating contribution of spin mechanisms in GaAs QWs
is also demonstrated for the circular MPGE~\cite{Lechner2011},  
where in-plane spin density required for spin-galvanic effect 
is created via optical excitation and the Hanle effect~\cite{PRB07,Nature02}.

Circular photogalvanic and spin galvanic effects have also been applied for
studying the SIA/BIA-ratio in  (001)-oriented doped InAs/InGaSb QWs and 
InGaAs/InAlAs QW structures~\cite{PRL04,PRB07}. 
In InAs/InGaSb QWs the measurements yield the value in the range  1.6 -- 2.3 which agrees 
well with theoretical results~\cite{Lommer88p728} predicting a dominating Rashba
spin-orbit coupling for InAs QWs and is also consistent with
experiments applying other transport methods~\cite{Knap96,Nitta97p1335}.
Note that the Rashba term is  very sensitive to details of the sample growth and further treatment. 
Furthermore, photogalvanic methods have been applied to study the SIA/BIA-ratio 
in a set of InGaAs/InAlAs QW structures with semitransparent gate~\cite{Kohda2012}. The 
measurements  supported by the weak antilocalization experiments 
permitted to find a proper QW design for the 
realization of the persistent helix conditions. 
These results are discussed in section~\ref{helixexp}.

To complete the picture we note that BIA and SIA induced CPGE, SGE and MPGE 
have been also observed in (001)-oriented InSb/(Al,In)Sb and HgTe/CdHgTe 
quantum well structures~\cite{Frazier2009KhodropPGE,Li2012,Stachel2012,Diehl2007MPGE001,Diehl2009}. 
These narrow band materials are of particular interest 
for spin physics because they are characterized by 
high mobility and small effective masses as well as by a very large $g^*$-factor and 
spin-orbit splitting~\cite{Khodaparast2004,Gilbertson2009,Kallaher2010,Akabori2008,Leontiadou2011}. 
So far, while confirmed the band spin splitting, 
most of the studies have been aimed to the mechanisms 
of the current formation in these novel materials, which can now be extended by 
special studies aimed to SIA/BIA interplay.

The experiments described above were carried out applying terahertz/microwave radiation. 
The dominant mechanism of the spin-orbit splitting, however, 
can also be determined from study of photogalvanics 
caused by  interband absorption~\cite{Golub_2003,Duc2010,Lu2011}. 
An interesting possibility to study the spin splitting provides
the study of the CPGE spectra~\cite{Golub_2003}: 
the SIA-induced CPGE photocurrent has a spectral sign inversion in contrast 
to the BIA-one. The interplay of the Rashba/Dresselhaus  has been 
investigated in applying CPGE in Ref.~\cite{Yu2012_1,Yu2011,Yu2012_2,Yu2013}
and MPGE in~\cite{Dai2010}.

To conclude this part the observation of the sign reversal of the 
Rashba/Dresselhaus-ratio upon changing  the $\delta$-doping position  
in the heterostructure together with quantum well width dependence of BIA can be used 
for growth of 2D structures with controllable spin splitting.
It is important to note that measurements have also been carried 
out at technologically important room temperature at which other methods based 
on spin-relaxation or antilocalization experiments can not be applied.
Thus, the measurements of photogalvanics can be used as a 
necessary feedback for technologists looking for perfectly symmetric 
structures with zero Rashba constant or for structures with equal 
Rashba and Dresselhaus spin-splittings. 
The latter will be discussed in the next section.

\subsection{Quantum well design requirements for long spin relaxation times in (001)-grown QWs
and realization of persistent spin helix
}
\label{helixexp}

The strongest anisotropy of the spin-orbit splitting can be achieved 
in zinc-blende semiconductor-based (001)-grown QWs with the 
$\bm k$-linear Rashba and Dresselhaus terms  of equal strength, $\alpha = \beta$.
Under these circumstances and for unessential contribution 
of $\bm k$-cubic BIA terms the spin splitting  vanishes in certain {$\bm k$}-space direction. 
Moreover, the resulting  effective magnetic field $\bm B_{eff}(\bm k)$ 
is aligned along one of the $\langle 110 \rangle$ crystallographic axes 
for any wavevector $\bm k$, see Fig.~\ref{A1spinfig01}g. 
Consequently, it becomes ineffective for  spins oriented along this axis. 
In this particular case, the interference of the Dresselhaus and Rashba terms leads 
to the i) disappearance of an anti-localization~\cite{Pikus95p16928,Knap96})
suppression of the Dyakonov-Perel  relaxation 
for spin oriented along $\bm B_{eff}$~\cite{Averkiev99PRB15582,Averkiev02pR271,omega4},  
iii) lack of SdH beating ~\cite{Tarasenko02p552,Averkiev2005} and iv) makes possible the formation of 
the persistent spin helix (PSH). The latter represents a new state of such a spin-orbit coupled system, 
which was predicted in Ref.~\cite{Bernevig2006} 
and experimentally observed in GaAs 2DES with weak $\bm k$-cubic Dresselhaus terms (see e.g.~\cite{Desrat2006})
applying transient spin-gating spectroscopy~\cite{Koralek2009,Walser2012Nature}. 
In this particular case spin precession around 
the fixed axis $\bm B_{eff} \parallel [1\bar{1}0]$ 
supports the space oscillations of the spin distribution  in the $[110]$ direction with a  
period $\pi\hbar^2/(2m^*\alpha)$. Indeed, the precession angle for electron spins aligned 
in the $(1\bar{1}0)$ plane equals to $2\pi$ after passing each period, while the spins 
oriented along $[1\bar{1}0]$ direction are intact at all. 
This demonstrates the stability of the space oscillating state (PSH state) 
to the spin precession. 
The specific spin splitting for $\alpha = \beta$ 
serving  novel ways for spin manipulation attracted valuable attention. 
There has been much effort in this field both
theoretically with new device proposals~\cite{Schliemann03p146801,JapaneseworkAPL2012}
and discussion of the PSH formation~\cite{Liu2006,Cheng2006,Bernevig2008,Li2010,Slipko,Slipko2013} as well as experimentally 
with the aim to obtain SIA equal to BIA~\cite{Koralek2009,Walser2012Nature,Scheid2007,Scheid2009,Kunihashi09,PRL04,LechnerAPL2009,Kohda2012}.

The design and growth of structures with a defined SIA/BIA-ratio 
needs techniques for its control. Generally, the requirement of 
$\alpha = \beta$ can be fulfilled   by the variation of 
both Rashba and Dresselhaus terms, which depend on  
a number of macroscopic parameters, such as  material of quantum well, 
quantum well width,  doping profile and growth temperature, 
gate voltage, carrier density, sample temperature etc.
The Dresselhaus SOI is primary determined by the material 
properties and quantum well width and 
is fixed for a given quantum well~\cite{doping_dress}.
%
Therefore, 
the only way to realize $\alpha = \beta$ 
in a given QW is to control the Rashba term. The latter can be achieved by the position of the 
asymmetric $\delta$-doping~\cite{Koralek2009,LechnerAPL2009,Kohda2012}, see section~\ref{001exp}, 
or by the application of a gate voltage~\cite{Kohda2012,Nitta97p1335,Engels97,Koga02,FanielKoga2011}. 
Figure~\ref{fig2apl}(a) shows that in $15$~nm wide GaAs QWs the $\alpha = \pm\beta$
condition is achieved for $\chi = 0$  and $\chi \approx 0.17$. 
In these structures the ratio of the BIA and SIA related 
photocurrents $|J_{x'}/J_{y'}|$ is about unity 
indicating that SIA and BIA have almost equal strengths.
Consequently, one obtains the spin splitting cancellation  either in $[1 \bar{1}0]$ or [110] 
crystallographic direction depending on the relative sign of the SIA and BIA terms.

While the weak $\bm k$-cubic SOI in GaAs based QWs barely affects the PSH formation, 
the important question arises whether a PSH type state will generally survive 
in materials with strong SOI where finite $\bm k$-cubic terms gain importance, 
in particular for heterostructures at higher charge carrier densities.
%
The effect of cubic in $\bm k$ terms on PSH 
has been analyzed in a few theoretical 
works~\cite{Pikus95p16928,Knap96,Kettemann2007,Glazov06,Stanescu07,Duckheim10,Luffe11} 
and has been demonstrated experimentally in Ref.~\cite{Kohda2012}.
%
The PSH conditions and the influence of the cubic in $\bm k$-terms on spin transport 
in a material with strong SOI have been studied in InGaAs quantum wells applying 
two complementary experiments, transport and photogalvanics~\cite{Kohda2012}. 
In this work strain-free (001)-grown ${\rm In_{0.53}Ga_{0.47}As / In_{0.52}Al_{0.48}As}$ 
quantum well structures hosting a two-dimensional electron gas were designed to achieve 
almost equal linear Rashba and Dresselhaus coefficients, $\alpha$ and $\beta$, at zero gate voltage. 
Since $\beta$ is usually much smaller than $\alpha$ in InGaAs 2DEGs~\cite{Faniel11}, one needed to 
enhance $\beta$ and to reduce the built-in Rashba SOI. The former condition
was achieved by making use the dependence of the Dresselhaus term on the QW widths,
$\beta \propto  1/L_w^2$, and growing sufficiently narrow QWs 
of width $L_w = 4${~nm}  and 7~nm. A small $\alpha$ at zero gate bias 
was obtained by preparing symmetric InGaAs QWs.  For that two {Si doping} 
layers with densities $n_1 = 1.2$  and $n_2 = 3.2 \times 10^{18}~{\rm cm}^{-3}$ 
were placed into the InAlAs barriers, each 6~nm away from the QW.
Here, the higher doping level on the top side of the QW compensates the surface charges. 
A fine tuning of the Rashba spin splitting was achieved by the gate voltage.

\begin{figure}
 \centering
  \includegraphics[width=1.0\columnwidth]{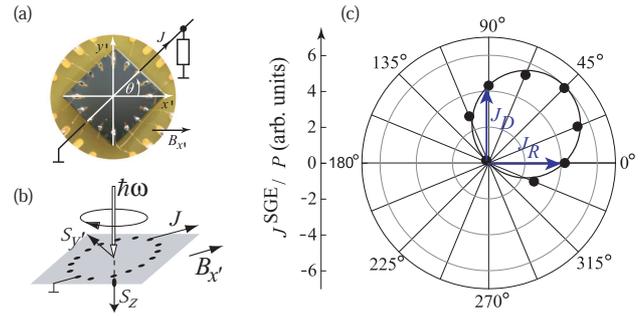}
\caption{
(a) Sample geometry and (b) sketch of the experimental arrangement  
used for measurements of the spin-galvanic effect in InGaAs quantum wells. 
For these measurements the samples  were irradiated by circularly polarized 
light along the growth direction, and an external magnetic field 
was applied along the $x'$-axis. 
The light generates a nonequilibrium spin polarization $\bm S \parallel z$ which, 
by means of the in-plane magnetic field, is rotated into the QW plane resulting in $S_{y'}$.
Such a non-equilibrium in-plane spin polarization causes a spin-galvanic effect~\cite{PRL04,Nature02}.
The photocurrent $J^{SGE}(\theta)$ is mapped by measuring successively signals from 
opposite contact pairs. (c)  Azimuthal dependence of the SGE current $J^{SGE}(\theta)$ 
measured in a 4~nm QW at room temperature, $\lambda = 148$~$\mu$m and at $B_x = 0.8$~T. 
The solid line shows the fit according to $J = J_{\rm R} \cos\theta + J_{\rm D} \sin\theta $ 
with the ratio of $ J_{\rm R} / J_{\rm D} = 0.98~\pm~0.08$. After~\cite{Kohda2012}.
%
}
\label{figure01_PSH}
\end{figure}

Figure~\ref{figure01_PSH} shows experimental geometry and 
the anisotropy of the spin-galvanic signal. The current is 
studied in 4~nm QW at room temperature applying radiation with wavelength 148~$\mu$m. 
It is measured along different in-plane directions determined by the 
azimuth angle $\theta$ with respect to the fixed in-plane magnetic field ${\bm B} \parallel x$.
The current component, $J_R$, parallel to the magnetic field is driven by the Rashba spin splitting, 
while the perpendicular component, $J_D$ is caused by the Dresselhaus SOI, 
see section~\ref{method}. 
The data can be well fitted by 
$J = J_{\rm R} \cos\theta + J_{\rm D} \sin\theta $, with $ J_{\rm R} / J_{\rm D} = 0.98 \pm 0.08$. 
This ratio is related to that between the linear Rashba and Dresselhaus SOI strengths, 
$ J_{\rm R} / J_{\rm D} = \alpha / \tilde{\beta}$. 
The renormalized coefficient $\tilde{\beta}$ is described by Eq.~\eqref{betatilde} and 
takes into account the influence of the first harmonic of the cubic in $\bm k$ 
spin-orbit terms on linear in $\bm k$ band spin splitting, see section~\ref{001}. 
The results of Fig.~\ref{figure01_PSH}(c) demonstrate that 
in ungated 4~nm QW samples the condition of
the PSH creation is fulfilled. 
The  $\alpha = \tilde{\beta}$ condition 
indicating the cancellation of the linear in $\bm k$ SOI has been also verified 
applying circular photogalvanic effect technique, see section~\ref{method}.
By contrast, for the 7~nm QW with smaller $\beta$ a substantially stronger SIA,  
$\alpha/\tilde{\beta} ~\approx  4$,  
has been measured applying both techniques. SGE and CPGE measurements 
carried out at low temperatures $T\approx 5$~K 
demonstrated a weak temperature dependence of $\alpha/\beta$.  The fact that the ratio of these spin-orbit constants is almost independent of temperature
in studied InAs quantum wells is in agreement with the theory.
Owing to a small electron effective mass (around 0.04$m_0$) and a high
electron density ($n_s = 3.5 \times 10^{12}$~cm$^{-2}$), the Fermi energy is about
170 meV. This means that the 2D electron gas is degenerate even at room
temperature. The temperature-dependent corrections to the Rashba and
Dresselhaus constants are in the order of the ratio of the thermal energy to
the Fermi energy which is less then 15~\% in the studied structure even at room temperature. 


\begin{figure}
 \centering
  \includegraphics[width=1.00\columnwidth]{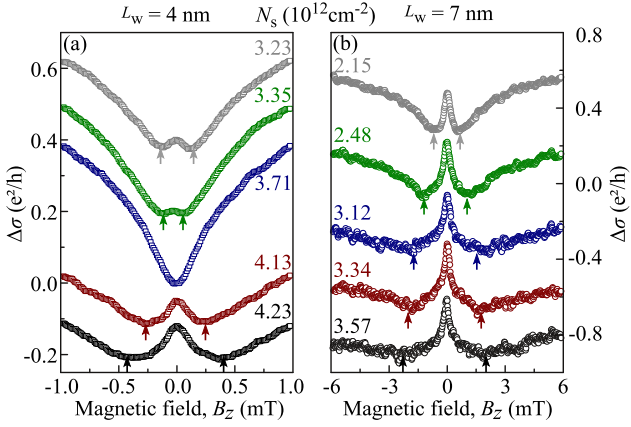}
\caption{
Magneto-conductance profiles (in units of $e^2/h$) measured at 
different gate voltages, i.e., carrier densities $N_s$, for (a) 
4~and  (b) 7~nm QWs InGaAs quantum wells
at $T =$ 1.4~K. 
All curves in (a) and (b) are shifted positively (gray and green) and negatively 
(black and red) with respect to the blue curve, for 
which $\Delta \sigma = 0$ at $B_z = 0$~mT.
For the 4~nm QW a clear WL dip occurs for a carrier density 
of $3.71 \times 10^{11}$~cm$^{-2}$, which is absent for the 7~nm QW. After~\cite{Kohda2012}.
}
\label{figure03}
\end{figure}

The SIA/BIA cancellation in 4~nm QWs has also been obtained in transport experiment 
where the quantum correction to the magneto-conductivity in the gated Hall bar structures 
was measured in the presence of an external magnetic field $\bm B$, pointing 
perpendicularly to the QW plane~\cite{Sch08}.
Figures~\ref{figure03}(a) and (b) show the measured magneto-conductance profiles at 
different gate voltages for the 4~nm and 7~nm wide QWs, respectively.
On the one hand, for the 7~nm QW, only weak antilocalization (WAL) characteristics are observed, which get 
enhanced with increasing $N_s$.
On the other hand, most notably, the magneto-conductance for the 4~nm QW near $B\!=\!0$ 
changes from WAL to weak localization (WL) characteristics and back again to WAL upon increasing $N_s$ 
from 3.23 to 4.23$\times 10^{12}$~cm$^{-2}$. 
The occurrence  of WL (at  $N_s \! = 3.71\times 10^{12}$~cm$^{-2}$) reflects suppressed 
spin relaxation, and the observed sequence WAL-WL-WAL unambiguously
indicates that -- even in presence of strong $\bm k$-cubic SOI -- a PSH 
condition is fulfilled in the WL region.

Comparison of the photocurrent measurements with the weak 
localization experiments enables us to extract information on 
the role of the cubic terms. Indeed, while the photocurrent experiments are 
insensitive to the third harmonic of the cubic term $\bm B_{eff}^{(3)}(\bm k)$,   
and thus reveal only the ratio $\alpha / \tilde{\beta}$, 
the transport experiment probes spin randomization due to the entire SOI contribution, 
Eqs.~\eqref{A1currentequ6},~\eqref{A1currentequ7} and~\eqref{HBIAcub}. 
A  numerical analysis of the WAL-WL-WAL transition 
applying $\alpha / \tilde{\beta}$ obtained from the photocurrent data 
clearly demonstrated that a PSH type state remains even for finite cubic SOI.
The essential prerequisite for this is that $\alpha$ and $\beta$ 
are close to each other, a condition which for InAs-based structures 
can be reached  in very narrow and almost symmetric QWs due 
to a specially designed doping profile.
However, in contrast to systems with dominating $\bm k$-linear 
spin splitting, the PSH is obtained for close, 
but nonequal Rashba and Dresselhaus strengths.

\subsection{Symmetry and spin dephasing in (110)-grown quantum wells}
\label{110exp}

Quantum well structures prepared on (110)-oriented GaAs substrates are 
of particular interest because in QWs of this orientation and special
design extraordinarily slow spin dephasing can be achieved. 
Spin lifetimes up to several nanoseconds~\cite{Ohno1999,Harley03,Dohrmann2004,Hall2005,PRL08110,Mueller2008,Mueller10,Hassenkam1997,Hall2003PRB,Henini2004,Morita2005,Ku2005,Hicks2006,Couto2007,Schreiber2007,Schreiber2007_2,Eldridge2010_4,Iba2010,Voelkl2011,Huebner2011}
or even submicroseconds~\cite{Griesbeck2012} have been reported in GaAs and other III-V semiconductor-based 
heterostructures, for  review see e.g.~\cite{Wu10}.
As discussed in section~\ref{110} in structures of this orientation the effective magnetic field induced by 
the bulk inversion asymmetry points along the growth 
axis and does not lead to the Dyakonov-Perel relaxation of spins 
oriented along this direction.  Therefore, in symmetrical (110)-grown QWs 
with SOI solely determined by BIA,  spin relaxation of the $z$-component is governed by   the 
Elliot-Yafet mechanism, see e.g.~\cite{Averkiev02pR271}, 
being rather ineffective in GaAs based QWs.
However, in asymmetric quantum wells this advantage  fades away due to the
Dyakonov-Perel spin relaxation caused by Rashba spin splitting.
In undoped samples this condition seems to be naturally fulfilled,  
as demonstrated by time- and polarization-resolved
transmission measurements in Ref.~\cite{Ohno1999}, where long spin 
lifetimes were found. However, in doped QW samples SIA is strongly 
affected by the impurity Coulomb potential and  growth of 
symmetrical QWs with negligible SIA becomes a challenging task. 
Discussion of this and other external factors limiting the spin relaxation 
time have been the subject of a large number of theoretical works, 
see e.g.~\cite{Nestoklon_2012,Review_Glazov_Sherman_Dugaev,Cartoixa2005_2,Wu2002,Chang2005,Tarasenko2009PRB80,Glazov2010,Zhou2010,Poshakinskiy2013}.

\begin{figure}
\includegraphics[width=0.95\linewidth]{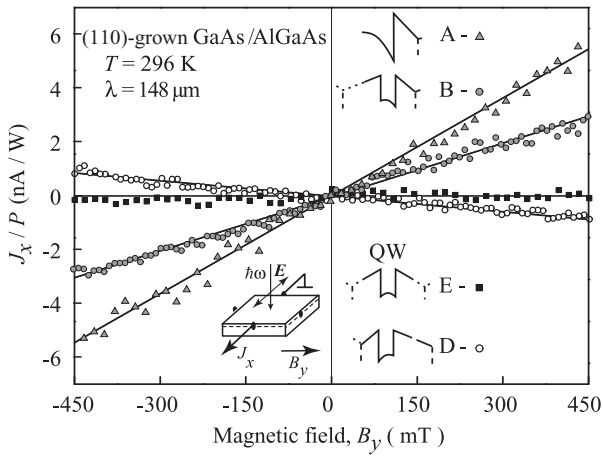}
\caption{Magnetic field dependences of the photocurrents measured in $x$-direction
for the radiation polarized along $x$ and the
in-plane magnetic field $\bm{B} \parallel y$.
The left inset   shows the experimental geometry. Four right insets show
the band profile and the $\delta$-doping position of the investigated samples.
After~\cite{PRL08110}.
}
\label{figure110exp}
\end{figure}

The degree of the structure inversion asymmetry has been analyzed 
in a set of double side $\delta$-doped QW samples with 
different parameter $\chi$ applying magneto-gyrotropic effect, 
see section~\ref{method}.
The degree of  SIA is reflected in the magnetic field dependence of the photocurrent 
displayed in Fig.~\ref{figure110exp}. From the symmetry arguments addressed in sections~\ref{110}
and~\ref{method} it
follows that for \textit{in-plane} magnetic field used for this measurements the 
MPGE current $J^{{\rm \: MPGE}}_x(B_y)$ is determined solely by SIA and, 
consequently, becomes possible only in the case of non-zero Rashba spin splitting. 
In line with these arguments we obtained that the slope of $J^{{\rm \: MPGE}}_x(B_y)$
reverses upon variation of the parameter $\chi$ from positive to negative values. 
As an important result of these measurements  the zero current response, i.e. zero SIA, is obtained 
for the almost symmetrically doped QWs, $\chi = 0$, grown at 480$^\circ$C~\cite{PRL08110,Olbrich2009_110}.
This is in contrast to (001)-oriented structures grown under standard conditions ($T > $600$^\circ$C)
where for QWs with $\chi = 0$ a substantial SIA is detected, see section~\ref{001exp} and Fig.~\ref{fig2apl}. 
This essential difference stems from the growth temperature, and,
subsequently, the impurity diffusion length. While for \textit{in-plane} magnetic field only 
SIA related MPGE is possible, for an \textit{out-of-plane} magnetic field $B_z$ the 
BIA related photocurrent is allowed and indeed observed for all samples. 
The latter demonstrate that for the structures with $\chi \neq 0$
SIA becomes important and, as demonstrated by complimentary TRKR experiments, spin relaxation 
accelerates. Note that studies of photocurrents 
excited in (110)-grown QW structures in the absence of an external 
magnetic field~\cite{Bieler05,Zhao2005,Shalygin2006,Diehl2007_110,Hu2008} are consistent with 
the results on MPGE and TRKR.

Study of spin relaxation in the  QW structure characterized by photogalvanic measurements 
and other double side $\delta$-doped QWs 
confirmed that in structures with symmetric doping ($\chi =0$)
the spin relaxation time is maximal. In these strutures the spin dephasing 
has been investigated applying  time- and polarization-resolved photoluminescence 
(TRPL)~\cite{Dohrmann2004,PRL08110,Olbrich2009_110}, 
spin noise spectroscopy~\cite{Roemer2007,Mueller2008,Mueller10}.
The measurements yield the record values of the spin 
dephasing times in GaAs up to  250~ns were obtained applying 
resonant spin amplification technique~\cite{Griesbeck2012}
and demonstrate that symmetrically doped (110)-oriented QW structures set 
the upper limit of spin dephasing in GaAs QWs.

\section{Interplay of BIA and SIA in other 2D systems}

\subsection{ Structure inversion asymmetry and spin splitting in wurtzite QWs}
\label{wurtziteexp}

\begin{figure}
\includegraphics[width=0.95\linewidth]{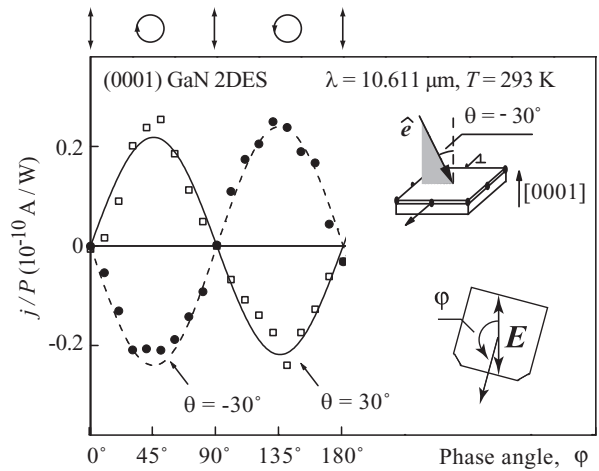}
\caption{Photocurrent in GaN QWs normalized by the radiation  power 
$P$ as a function of the phase angle $\varphi$ 
defining helicity. Measurements are presented for room 
temperature and irradiation by light of Q-switched CO$_2$ laser at the 
wavelength  $\lambda$ = 10.61~$\mu$m. 
The current $j_x$ is measured for direction perpendicular to propagation of light 
(angle of incidence  $\Theta_0$ = 30$^\circ$ ). 
Solid and dashed lines show calculated CPGE photocurrent. 
Insets sketch the experimental geometry and rotation of 
the $\lambda/4$-polarized  by the angle $\varphi$ in respect to linearly
polarized laser radiation field, $\bm E$.
The ellipses on top of the panel illustrate the 
polarization states for several angles $\varphi$.
After~\cite{WeberAPL2005GaN}.
}
\label{figure06GaN}
\end{figure}

Wurtzite low-dimensional structures, in particular wide bandgap GaN, 
has been extensively investigated for  applications as
blue and ultraviolet light sources~\cite{Nakamurabook} as well as for  high temperature
and high power electronic devices.~\cite{GaN1,GaN2,GaN3} The commercial fabrication of
blue and green LEDs has led to well established technological
procedures of epitaxial GaN preparation and sparked a great research
activity on the properties of heterostructures based
on GaN and its alloys with AlN and InN. Two-dimensional 
GaN also attracted growing attention as a potentially interesting 
material system for spin physics since, doped with manganese, 
it is expected to become ferromagnetic with a Curie-temperature above
room temperature~\cite{Dietl2000}; 
being gadolinium doped it may offer an opportunity 
for fabricating magnetic semiconductors~\cite{7gado,7gadobis,8gado,9gado,Buss2013Gd}; and
GaN-based structures show rather long spin relaxation times~\cite{Beschoten01,Buss2010,Buss2011}.
A further important issue is the existence of considerable Rashba spin-splitting 
in the band structure. 
First indications of substantial spin-orbit splitting
came from the observation of the SIA-type circular photogalvanic effect in GaN 
heterojunctions at Drude absorption of THz radiation~\cite{WeberAPL2005GaN}. 
Figure~\ref{figure06GaN} shows the photocurrent as a function of the phase angle $\varphi$
defining the radiation helicity. A finger print of the CPGE - reversing of the current direction
upon switching helicity from right to left handed circularly polarized light - is clearly detected. 
The observed CPGE current always flows perpendicular to the incidence plane 
and its magnitude does not change upon rotation of the in-plane component 
of the light propagation unit vector $\hat {\bm e}$. 
The reason of this axial isotropy is that in wurtzite type structures 
both, SIA and BIA, lead to the same form of spin-orbit interaction given by Eq.~\eqref{wurtzit}, 
see section~\ref{sectionwurtzite}.
Therefore, they cause the linear coupling  between orthogonal 
vectors  (here photocurrent $\bm j$ and pseudovector $P_{\rm circ} \hat {\bm e}$) only.
The band spin-splitting, which is actually not expected in
wide band-gap semiconductors, in GaN/AlGaN heterostructures 
is caused by a large piezoelectric effect~\cite{CingolaniPRB2000} 
yielding a strong electric field at the GaN/AlGaN interface.
This electric field causes  a polarization induced doping effect~\cite{Litvinov}, and,
on the other hand, results in a sizable Rashba contribution to 
the band spin-splitting.
Making use of intraband, intersubband and interband absorption, 
the investigations of photogalvanic phenomena were extended to 
GaN quantum wells of various design 
as well as to low dimensional wurtzite structures under 
uniaxial strain, demonstrating the Rashba character of the 
spin splitting~\cite{Tang2007GaN,Zhao07GaN,PRB2008,Wittmann2008,Cho2007_1,WeberAPL2005GaN,HeAPL2007,TangApl2007,ChoPRB2007,SSC07,Tang2008}.
The  values of the spin splitting, up to 1~meV 
at the Fermi wavevector, have been obtained by magneto-transport 
measurements~\cite{ChoApl2005,ThillosenAPL2006,SchmultPRB2006,TangAPL2006,Spirito2011}.
This results also confirmed that spin splitting is dominated by 
the $\bm k$-linear terms and the $\bm k$-cubic contribution 
is negligible.
%
%
%
Note that studies on photogalvanic effects in 
InN- and ZnO- heterostructures demonstrated substantial band spin splitting 
also in these wurtzite materials~\cite{Zhang2010ZnO,Zhang2008InN,Zhang2009InN_1,Zhang2009InN,Duan2013_ZnO}. 

\subsection{Structure inversion asymmetry and spin splitting in SiGe QWs}
\label{SiGEexp}

\begin{figure}
\includegraphics[width=0.95\linewidth]{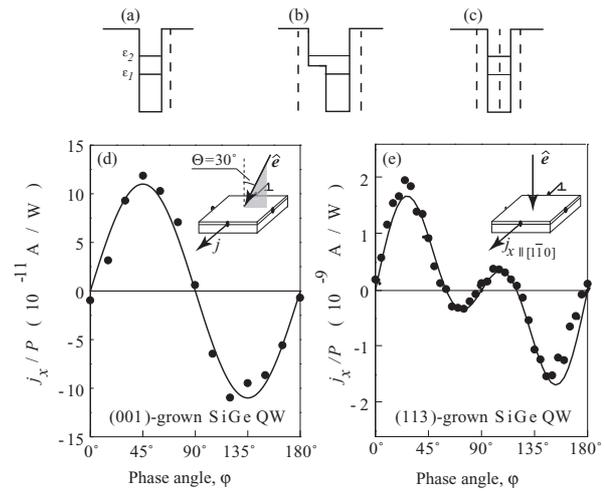}
\caption{Potentials profiles of investigated samples:
(a) asymmetrically  doped compositionally symmetric QW, (b) 
compositionally stepped QW, and (c) symmetric QW. 
The vertical dashed lines indicate the doping.
(d-e) Photogalvanic current $j_x$ normalized by the light power $P$ 
and measured at room temperature as a function of the phase angle $\varphi$.
(d) detected in (001)-grown and asymmetrically doped
SiGe QWs. The data were obtained under oblique incidence $\Theta_0$= 30$^\circ$
of irradiation at $\lambda = 10.6\,\mu$m.
The full line is fit to theory. 
(e) detected in (113)-grown SiGe QWs.
The results were obtained under normal incidence
of irradiation at  $\lambda = 280\,\mu$m at room temperature.
The full line is fit to theory. 
The insets in (d) and (e) show the corresponding experimental geometries.
The ellipses on top of the panel illustrate the 
polarization states for several angles $\varphi$. After~\cite{A1PRB02SiGe}.
}
\label{figureSiGEexp}
\end{figure}

Experimental evidence of the spin degeneracy removal was in focus of the first work on
photogalvanics in SiGe quantum wells~\cite{A1PRB02SiGe}. 
Experiments on doped structures 
of various design demonstrated that SIA is the necessary prerequisite 
for the band spin splitting and  generation of CPGE. 
SIA is obtained by asymmetric doping, see Fig.~\ref{figureSiGEexp}(a), and/or using of 
stepped potential, see Fig.~\ref{figureSiGEexp}(b). 
Circular photogalvanic effect is detected for both systems but was absent in the symmetric QWs 
depicted in Fig.~\ref{figureSiGEexp}(c). 
Examples of the photocurrent's helicity dependences  are shown in Fig.~\ref{figureSiGEexp}(d) 
and (e) for structures grown on (001)- and (113)-oriented substrates, 
respectively. The measurements confirm that for (001)-grown
QWs CPGE is generated only for oblique incident circularly polarized 
light which provides an in-plane component of the photon angular momentum. 
The CPGE current flows normal to the plane of incidence and for a fixed angle of incidence 
its strengths remains constant for any light propagation direction. 
As both CPGE and band spin splitting are described by the equivalent 
second rank pseudo-tensors, see section~\ref{method}, this observation supports the conclusion 
that the band spin splitting in these structures is given by 
the Rashba term~Eq.~\eqref{A1currentequ7}.
The appearance of the CPGE in the stepped QWs does not 
contradict with the results of Eldridge~et~al.~\cite{Eldridge2011}
demonstrating that in undoped 2D structures 
Rashba coefficient can be negligibly small despite huge conduction-band 
potential gradients which break the inversion symmetry.
In the discussed case we deal with doped structures for which
an asymmetric shape  of QWs results in asymmetry of the dopant Coulomb force acting 
on free electrons. In line with symmetry arguments for symmetrically
doped rectangle QW no photogalvanic currents have been detected. 
A substantial SIA  in asymmetrically doped SiGe QWs has been also 
confirmed by experiments on electron spin 
resonance~\cite{A1Wilamowski02p195315,Malissa2004,Tyryshkin2005,Matsunami2006} and 
magneto-gyrotropic photogalvanic effect~\cite{PureSpin2007} and 
CPGE at interband absorption~\cite{Wei2007}.

In (113)-grown SiGe QWs the photocurrent mostly comes from the normal incidence, 
see Fig.~\ref{figureSiGEexp}(e)~\cite{A1PRB02SiGe}. The reason for the CPGE current
excited by normally incident light is the reduction of symmetry from C$_{2v}$ to C$_s$
and the arguments are the same as that used for discussion of band spin splitting in 
(113)-oriented QWs, see section~\ref{113}. Observation of such CPGE
current with magnitudes comparable to that detected in GaAs QW structures 
indicates appearance of the  band spin splitting for spins oriented 
normal to QW plane. We note that the large circular photogalvanic effect
in (113)-grown SiGe QWs was not only applied for studying 
SIA/BIA interplay but has also been used for development of 
the all-electric detector of light Stokes parameters~\cite{Detector2007}.

\subsection{SIA/BIA interplay in (113)-, (112)-, (013)-oriented and miscut (001)-grown  
zinc-blende structure based 2DES and artificial symmetry reduction}
\label{sia113}

In the last part of the review we briefly address the results 
obtained for zinc-blende structure based quantum wells of 
less spread crystallographic orientation.  A specific property of 
(113)-, (112)-, (013)-oriented and miscut (001)-grown zinc-blende structure based 2DES
is the presence of spin splitting for spins oriented  
normal to the quantum well plane, see section~\ref{113}. 
These systems, apart from (013) oriented QWs, 
belong to C$_s$ point group and have a mirror reflection
plane $m_1$ normal to the QW plane, i.e. similar to the plane $m_1$ 
in asymmetric (110)-grown QWs depicted in Fig.~\ref{figure01}(c). 
The reduction of symmetry gives rise to 
circular photogalvanic effect at normal incidence
and spin-galvanic effect for $\bm S \parallel z$,  
which are forbidden for (001)-oriented QWs. 
These effects have been detected in miscut MOCVD (001)-grown GaAs QWs~\cite{GanichevPrettl,PRL01,GanichevAPL2000} as well as in  
(113)-grown GaAs-, SiGe-based two dimensional 
structures~\cite{PRL2002bleching,A1PRB02SiGe,Ganichev02,Vasyukov2010,Vasyukov2013}
yielding a helicity dependent photocurrent in the direction $x$
perpendicular to $m_1$. The observation of the normally incident light induced photocurrent 
reflects the band spin splitting for  electrons moving along $x$ direction
and allows to determine large $g^*$-factor of holes~\cite{Vasyukov2010,Vasyukov2013}.

While the circular photogalvanic effect at normal incidence has also 
been observed for (013)-oriented HgTe QWs the in-plane photocurrent  
direction  changes depending on various macroscopic 
parameters, e.g. temperature~\cite{Wittmann2010}. 
This observation indicates that photocurrents as well as 
the band spin-splitting addressed in section~\ref{method}, may arbitrary 
change upon variation of QW design and experimental conditions. 
The reason for this behaviour is that (013)-oriented  quantum wells 
belong to the trivial point group C$_1$ lacking any symmetry operation except 
the identity.  Hence, no preferential direction of the
circular photocurrent  or band spin splitting is forced
by the symmetry arguments, see section~\ref{113}. It is important to note,
that owing to strong spin-orbit coupling in HgTe-based QWs
the CPGE has been observed to be about an two orders of magnitude larger than 
that in GaAs, InAs and SiGe low dimensional structures.
The large helicity-dependent photoresponse obtained in the wide range
of radiation frequencies suggests HgTe QW structures
are promising  for detection of THz/IR radiation, in particularly,
for the all-electric detection of the radiation Stokes
parameters~\cite{Danilov2009}.

Finally we note that the symmetry reduction and, consequently, 
band spin splitting can be obtained applying strain even to bulk materials
or depositing asymmetric lateral structures on the top of quantum well.
The former way has been successfully used to obtain spin polarization 
by electric current (inverse spin galvanic effect)~\cite{inversedSGE,Ganichev04p0403641,Silov2004p5929,Sih2005,Yang06p186605}, 
in (001)-grown bulk InGaAs layers and GaAs membranes~\cite{Kato2004p176601}.
Strain has also been used to study photogalvanics and Rashba spin splitting 
in zinc-blende structures~\cite{Bieler05} and wurtzit GaN-based 2DES~\cite{HeAPL2007}.
The measurements indicate a substantial $\bm k$-linear 
band spin splitting which is forbidden without strain in 
these bulk materials.  The possibility of artificial symmetry reduction and variation of BIA and SIA  has 
also been demonstrated  by photogalvanic studies of asymmetric lateral 
superlattices~\cite{kotthaus,Olbrich2011Ratchet,Olbrich2009Ratchet,Kannan2011,IvchenkoRatchet}
and of structures with periodic quasi-one dimensional  wires~\cite{Jiang2011_1,Jiang2011_2}.

\section{Conclusions and outlook}
\label{conclusions}

Physics of momentum dependent Rashba/Dresselhaus  splitting of spin subbands in 
two-dimensional condensed matter systems has already resulted in 
a great variety of fascinating effects. 
This relativistic phenomenon caused by combined effect of atomic spin-orbit coupling and structure or bulk 
inversion asymmetry becomes possible in gyrotropic class of crystals 
and its form and strength can  be strongly affected by the interplay of Rashba and Dresselhaus effects. 
The key issue in this interplay is the point group symmetry 
allowing for certain crystallographic configurations and 
structure design cancellation of the BIA and SIA  
or separation of Rashba and Dresselhaus band spin splitting. Several 
specific configurations may give rise to the extraordinary long spin 
relaxation states or persistent spin helix. An access to analysis
of the SIA/BIA anisotropy even at technologically important room temperature
provides investigation of several types of photocurrents 
belonging to the class of photogalvanic effects.
These studies have been already used to demonstrate a possibility
of the  controllable variation of SIA by means of asymmetric delta-doping; 
to design (110)-grown QWs showing record spin relaxation times and (001)-oriented QWs
with fulfilled spin helix state condition; 
to explore the role of segregation and crystallographic orientation in the SIA/BIA 
strength and anisotropy; 
resulted in observation of SIA/BIA in wurtzite materials and SiGe QWs
and have been applied to study exchange interaction in diluted 
magnetic QWs~\cite{Ganichev2009DMS,Terentev2011,Olbrich2012}. 
The fact that photogalvanic effects are very general and have been detected in a large number 
of various 2DES makes them a proper tool in the 
arsenal of methods sensitive to subtle details 
of spin orbit interaction.  Particularly prospective for the further studies seems to be 
the contactless determination of the photogalvanic current anisotropy 
by the terahertz time-domain spectroscopy based experiments. Finally, 
we anticipate, that the interplay of Rashba and Dresselhaus effects 
will continue to be a manifold important tool in spin physics of low-dimensional systems 
giving rise to many new exciting phenomena.

\begin{acknowledgement}
We are grateful to S.~N.~Danilov and C. Zoth for valuable discussions.
This work was supported by DFG  (SPP~1285), Linkage Grant of IB of BMBF at DLR, and RFBR.
\end{acknowledgement}


\begin{thebibliography}{99}


\bibitem{A1Dresselhaus55p580} G.~Dresselhaus, 
Phys. Rev. \textbf{ 100}, 580 (1955).

\bibitem{A1Dyakonov86p110} M.\,I.~D'yakonov and V.\,Yu.~Kachorovskii,
Fiz. Tekh. Poluprovodn. \textbf{20}, 178 (1986) [Sov. Phys.
Semicond. \textbf{ 20}, 110 (1986)].

\bibitem{A1Rashba60p1109} E.\,I.~Rashba, 
Fiz. Tverd. Tela \textbf{ 2}, 1224 (1960) [Sov. Phys. Sol.
State \textbf{ 2}, 1109 (1960)]. 

\bibitem{A1Bychkov84p78} Y.\,A.~Bychkov and E.\,I.~Rashba,
Pis'ma Zh. \`{E}ksp. Teor. Fiz. \textbf{ 39}, 66 (1984) [JETP
Lett. \textbf{ 39}, 78 (1984)].

\bibitem{Krebs1996} O. Krebs and P. Voisin, Phys. Rev. Lett. \textbf{77}, 1829 (1997).

\bibitem{Krebs1997} O. Krebs, W. Seidl, J.P. Andre, D. Bertho, C. Jonani and P. Voisin, Semicond. Sci. Technol. 12, 938 (1997).

\bibitem{A1Vervoort97p12744} L.~Vervoort  and P.~Voisin,
Phys. Rev. B \textbf{ 56}, 12744 (1997).




\bibitem{spintronicbook02} I.~Zutic, J.~Fabian, and S.~Das-Sarma,
Rev. Modern Phys. \textbf{76}, 323 (2004).

\bibitem{Winkler2003} R. Winkler, Spin-orbit Coupling Effects in Two-Dimensional Electron
and Hole Systems (Springer, Berlin, 2003).

\bibitem{Maekawa} S. Maekawa,
Concepts in Spin Electronics (Oxford University Press, Oxford,
2006).

\bibitem{Fabian08} J. Fabian, A. Matos-Abiague, C. Ertler, P. Stano, and I. Zutic,
Acta Physica Slovaca \textbf{57}, 565 (2007).

\bibitem{Dyakonovbook} M.\,I.~Dyakonov (ed.), Spin Physics in Semiconductors,
(Springer, Berlin, 2008).

\bibitem{Awschalombook2008} T. Dietl, D.\,D. Awschalom, and M. Kaminska
(eds.), Spintronics (Semiconductors and Semimetals) (Academic
Press Inc., London, 2008).

\bibitem{Awschalombook2009}  D.\,D. Awschalom, R.\,A. Buhrman,
J.\,M.~Daughton, S.~von~Molnar, and M.\,L.~Roukes (eds.), Spin
Electronics (Kluwer Acad. Publ., Dordrecht, 2009).

\bibitem{Awschalombook2010} D.\,D. Awschalom, D. Loss, and N.~Samarth (eds.),
Semiconductor Spintronics and Quantum Computation
(Springer, Berlin, 2010).

\bibitem{Wu10} M.\,W. Wu, J.\,H. Jiang, and M.\,Q.~Weng,
Phys. Reports \textbf{493}, 61 (2010).

\bibitem{HandbookZutic} \textit{Handbook of Spin Transport and Magnetism}
eds. E.Y. Tsymbal and I. Zutic  (Chapman and Hall 2011).

\bibitem{SemicSpintronics} J.~Xia, W.~Ge, and K.~Chang, Semiconductor Spintronics (World Scientific, Singapore, 2012).

\bibitem{SpinCurrent} S.~Maekawa, S.\,O.~Valenzuela, E.~Saitoh, and T.~Kimura (eds.), 
Spin Current (Oxford University Press, Oxford, 2012).



\bibitem{DP71} M.~I. D'yakonov and V.~I.~Perel',
Sov. Phys. Solid State {\bf 13}, 3023 (1972).


\bibitem{Averkiev99PRB15582}  N.\,S.~Averkiev and L.\,E.~Golub,
Phys. Rev. B {\bf 60}, 15582 (1999);


\bibitem{Averkiev02pR271}  N.\,S.~Averkiev,
L.\,E.~Golub, and M.~Willander,
J.~Phys.: Condens. Matter {\bf 14}, R271 (2002).

\bibitem{omega4} N.S. Averkiev and L.E. Golub, Semicond. Sci. Technol. {\bf 23}, 114002 (2008).

\bibitem{Schliemann03p146801} J.~Schliemann, J.\,C.~Egues, and D.~Loss,
Phys. Rev. Lett. {\bf 90}, 146801 (2003).

\bibitem{Bernevig2006} B.\,A. Bernevig, B.\,A.~Orenstein, and S.-C.~Zhang,
Phys. Rev. Lett. \textbf{97}, 236601 (2006).

\bibitem{Koralek2009} J.\,D.~Koralek, C.\,P.~Weber, J.~Orenstein, B.\,A.~Bernevig,
Shou-Cheng Zhang, S.~Mack, and D.\,D.~Awschalom,
Nature \textbf{458}, 610 (2009).

\bibitem{Walser2012Nature} 
M. P. Walser, C. Reichl, W. Wegscheider, and G. Salis,
Nature Phys. \textbf{8}, 757 (2012).

\bibitem{review2003spin} S.\,D.~Ganichev  and W.~Prettl,
J. Phys.: Condens. Matter {\bf 15}, R935 (2003).




\bibitem{Ohno1999} 
Y.~Ohno, R.~Terauchi, T.~Adachi, F.~Matsukura, and H.~Ohno, Phys.
Rev. Lett. \textbf{83}, 4196 (1999).

\bibitem{Harley03} O. Z. Karimov, G. H. John, R.T. Harley,
W. H. Lau, M. E. Flatte, M. Henini, and R. Airey,
Phys. Rev. Lett. \textbf{91}, 246601 (2003).

\bibitem{Dohrmann2004} S.~D{\"o}hrmann 
D. H{\"a}gele, J. Rudolph, M. Bichler, D. Schuh, and M. Oestreich,
Phys. Rev. Lett. \textbf{93}, 147405 (2004).
%


\bibitem{Hall2005} K.~C. Hall, K.~G{\"u}ndo\v{g}du, J. L. Hicks, A. N. Kocbay, 
M. E. Flatte, T. F. Boggess, K. Holabird, A. Hunter, D. H. Chow,
and J. J. Zinck, 
Appl. Phys. Lett. \textbf{86}, 202114 (2005).
%

\bibitem{PRL08110}  V.\,V. Bel'kov, P.~Olbrich, S.\,A.~Tarasenko, D.~Schuh, W.~Wegscheider,
T.~Korn, C.~Sch{\"u}ller, D.~Weiss, W.~Prettl, and
S.\,D.~Ganichev, 
Phys. Rev. Lett. \textbf{100}, 176806 (2008).

\bibitem{Roemer2007} M.~R{\"o}mer, J.~H{\"u}bner, and M.~Oestreich, Rev. Scient. Instr.
\textbf{78}, 103903 (2007).

\bibitem{Mueller2008} 
G.\,M.~M{\"u}ller, M.~R{\"o}mer, D.~Schuh, W.~Wegscheider,
J.~H{\"u}bner, and M.~Oestreich, Phys. Rev. Lett. \textbf{101},
206601 (2008).



\bibitem{Mueller10} G.\,M.~M{\"u}ller, M.~Oestreich, M.~R{\"o}mer, and J.~H{\"u}bner,
Physica E \textbf{43}, 569 (2010).


\bibitem{Balocchi2011} A. Balocchi, Q. H. Duong, P. Renucci, B. L. Liu, C. Fontaine, T. Amand, D. Lagarde, and X. Marie, Phys. Rev. Lett. \textbf{107}, 136604 (2011).

%
\bibitem{Griesbeck2012} M.~Griesbeck, M.\,M.~Glazov, E.\,Ya.~Sherman, D.~Schuh,
W.~Wegscheider, C.~Sch{\"u}ller, and T.~Korn, Phys. Rev. B
\textbf{85}, 085313 (2012). 



\bibitem{Ye2012} H. Q. Ye, G. Wang, B. L. Liu, Z. W. Shi, W. X. Wang, C. Fontaine, A. Balocchi, T. Amand, D. Lagarde, P. Renucci, and X. Marie, Appl. Phys. Lett \textbf{101}, 032104 (2012).

\bibitem{Biermann2012} K. Biermann, A. Hern\'{a}ndez-M\'{i}nguez, R. Hey, and P. V. Santos, J. Appl. Phys. \textbf{112}, 083913 (2012).

\bibitem{Hernandez2012} 
Author(s): A. Hern\'{a}ndez-M\'{i}nguez, K. Biermann, R. Hey, and P. V. Santos, 
Phys. Rev. Lett. 109, 266602 (2012).

\bibitem{Wang2013} G. Wang, A. Balocchi, D. Lagarde, C. R. Zhu, T. Amand,  P. Renucci,  Z. W. Shi, W. X. Wang, B. L. Liu, and X. Marie, Appl. Phys. Lett \textbf{102}, 242408 (2013).



\bibitem{Cartoixa2003} X. Cartoixa, D.\,Z.-Y.~Ting, and Y.-C.~Chang,
Appl. Phys. Lett. \textbf{83}, 1462 (2003).


\bibitem{Hall2003APL} 
K.\,C.~Hall, W.\,H.~Lau, K.~G{\"u}ndo\v{g}du, M.\,E.~Flatt\'{e},
and T.\,F.~Boggess, Appl. Phys. Lett. \textbf{83}, 2937 (2003).

\bibitem{Cartoixa2005} 
X.~Cartoixa, D.\,Z.\,Y.~Ting, Y.\,C.~Chang, J. of Supercond.
\textbf{18}, 163  (2005).

\bibitem{JapaneseworkAPL2012} 
Y. Kunihashi, M. Kohda, H. Sanada, H. Gotoh, T. Sogawa, and J. Nitta,
Appl. Phys. Lett. \textbf{100}, 113502   (2012).




\bibitem{Jusserand92} B.~Jusserand, D.~Richards, H. Peric, and B.~Etienne,
Phys. Rev. Lett. \textbf{69}, 848 (1992).

\bibitem{Jusserand95p4707} B.~Jusserand, D.~Richards, G.~Allan, C.~Priester,  and B.~Etienne,
Phys.~Rev.~B {\bf 51}, 4707 (1995).


\bibitem{Pikus95p16928} F.\,G.~Pikus and G.\,E.~Pikus,
Phys. Rev. B. {\bf 51}, 16928 (1995).

\bibitem{Knap96} W. Knap, C.~Skierbiszewski, A.~Zduniak, E.~Litwin-Staszewska,
D.~Bertho, F.~Kobbi,  J.\,L.~Robert, G.\,E.~Pikus, F.\,G.~Pikus,
S.\,V.~Iordanskii, V.~Mosser, K.~Zekentes, and
Yu.\,B.~Lyanda-Geller, 
Phys. Rev. B \textbf{53}, 3912 (1996).

\bibitem{Miller03} J.\,B.~Miller, D.\,M.~Zumb{\"u}hl, C.\,M.~Marcus, Y.\,B.~Lyanda-Geller,
D.~Goldhaber-Gordon, K.~Campman, and A.\,C.~Gossard,
Phys. Rev. Lett. \textbf{90}, 076807 (2003).


\bibitem{Yu2008} G. Yu, N. Dai, J. H. Chu, P. J. Poole and S. A. Studenikin,
  {Phys. Rev. B} {\bf 78}, {035304} (2008). 



\bibitem{Glazov2009} M.\,M.~Glazov and L.\,E.~Golub,
Semicond. Sci. Technol. \textbf{24}, 064007 (2009).



\bibitem{Minkov2004} G.M. Minkov, A.V. Germanenko, O.E. Rut, A.A. Sherstobitov, L.E. Golub, B.N. Zvonkov and M. Willander, Phys. Rev. B {\bf 70}, 155323 (2004).

\bibitem{Scheid2007} M. Scheid, M. Kohda, Y. Kunihashi, K. Richter, and J.~Nitta,
Phys. Rev. Lett. \textbf{101}, 266401 (2008).

\bibitem{Scheid2009} M. Scheid, I. Adagideli, J. Nitta, and K.~Richter,
Semicond. Sci. Tech. \textbf{24}, 064005 (2009).

\bibitem{Kunihashi09} Y. Kunihashi, M. Kohda, and J. Nitta,
Phys. Rev. Lett. \textbf{102}, 226601 (2009).

\bibitem{PRL04} S.\,D.~Ganichev, V.\,V.~Bel'kov, L.\,E.~Golub, E.\,L.~Ivchenko, Petra~Schneider,
S.~Giglberger, J.~Eroms, J.\,De~Boeck, G.~Borghs, W.~Wegscheider,
D.~Weiss, and W.~Prettl,
Phys. Rev. Lett. \textbf{92}, 256601 (2004).

\bibitem{PRB07} S.~Giglberger, L.\,E.~Golub,  V.\,V.~Bel'kov, S.\,N.~Danilov, D.~Schuh, Ch.~Gerl,
F.~Rohlfing,  J.~Stahl, W.~Wegscheider, D.~Weiss, W.~Prettl, and
S.\,D.~Ganichev, 
Phys. Rev. B \textbf{75}, 035327 (2007).

\bibitem{LechnerAPL2009}  V.~Lechner, L.\,E.~Golub, P.~Olbrich, S.~Stachel, D.~Schuh,
W.~Wegscheider, V.\,V.~Bel'kov, and S.\,D.~Ganichev,
Appl. Phys. Lett. \textbf{94}, 242109 (2009).




\bibitem{Bieler05}   M.~Bieler, N.~Laman, H.\,M.~van~Driel, and A.\,L.~Smirl,
Appl. Phys. Lett. \textbf{86}, 061102 (2005).

\bibitem{Yang06p186605} C.L.~Yang, H.T.~He, Lu~Ding, L.J.~Cui, Y.P.~Zeng, J.N.~Wang, and W.K.~Ge,
Phys. Rev. Lett. \textbf{96}, 186605 (2006).

\bibitem{Tang2007GaN} Y. O. Tang, B.~Shen, H. W. He, N.~Tang, W. H.c Chen,
Z. J. Yang, G. Y.~Zhang, Y. Hc Chen, C. G. Tang, Z.\,G.~Wang,
K. S. Cho, and Y. F. Chen, Appl. Phys. Lett. \textbf{91}, 071920 (2007).

\bibitem{Zhao07GaN} H. Zhao, B. Liu, L.~Guo, C.~Tan, H.~Chen, and D.~Chen,
Appl. Phys. Lett. \textbf{91}, 252105 (2007).

\bibitem{Frazier2009KhodropPGE} M. Frazier, J. A. Waugh, J. J. Heremans, M. B. Santos, X. Liu, and
G. A. Khodaparast, J. Appl. Phys. \textbf{106}, 103513 (2009).

\bibitem{Yu2012_1} J.\,L.~Yu, Y.\,H.~Chen, Y.~Liu, C.\,Y.~Jiang,
H.~Ma, and L\,P.~Zhu, Appl. Phys. Lett. \textbf{100}, 152110 (2012).

\bibitem{GolubKochereshko} N.\,S.~Averkiev, L.\,E.~Golub, A.\,S.~Gurevich, V.\,P.~Evtikhiev,
V.\,P.~Kochereshko, A.\,V.~Platonov, A.\,S.~Shkolnik, and
Yu.\,P.~Efimov, Phys. Rev. B \textbf{74}, 033305 (2006).

\bibitem{Eldridge08} P.\,S.~Eldridge, W.\,J.\,H.~Leyland, P.\,G.~Lagoudakis, O.\,Z.~Karimov, M.~Henini, D.~Taylor, R.\,T.~Phillips, and
R.\,T.~Harley,
Phys. Rev. B \textbf{77}, 125344 (2008).

\bibitem{Larionov2008} A.\,V.~Larionov and L.\,E.~Golub, Phys. Rev. B
\textbf{78}, 033302 (2008).


\bibitem{Eldridge2010_2} P.\,S.~Eldridge, W.\,J.\,H.~Leyland, P.\,G.~Lagoudakis, R.\,T.~Harley, R.\,T.~Phillips, R.~Winkler, M.~Henini, and
D.~Taylor,
Phys. Rev. B \textbf{82}, 045317 (2010).


\bibitem{Eldridge2011} P.\,S.~Eldridge, J.~H{\"u}bner, S.~Oertel,
R.\,T.~Harley, M.~Henini, and M.~Oestreich, 
Phys. Rev. B \textbf{83}, 041301 (2011).
%


\bibitem{Stich07} D.~Stich, J.~Zhou, T.~Korn, R.~Schulz, D.~Schuh, W.~Wegscheider, M.\,W.~Wu, and C.~Sch{\"u}ller,
Phys. Rev.  Lett. \textbf{98}, 176401 (2007).

\bibitem{Cheng08} J.\,L.~Cheng, M.\,W.~Wu, and I.\,C.~da~Cunha~Lima,
Phys. Rev. B \textbf{75}, 205328 (2007).

\bibitem{Stich07_2} D.~Stich, J.\,H.~Jiang, T.~Korn, R.~Schulz, D.~Schuh, W.~Wegscheider, M.\,W.~Wu, and C.~Sch{\"u}ller,
Phys. Rev. B \textbf{76}, 073309 (2007).

\bibitem{Meier07} L.~Meier, G.~Salis, I.~Shorubalko, E.~Gini, S.~Sch{\"o}n, and K.~Ensslin,
Nature Physics \textbf{3}, 640 (2007).

\bibitem{Korn08} T. Korn. D. Stich, R. Schulz, D. Schuh, W. Wegscheider, and C. Sch{\"u}ller,
Physica E \textbf{40}, 1542 (2008).

\bibitem{Meier08} L.~Meier, G.~Salis, E.~Gini, I.~Shorubalko, and K.~Ensslin,
Phys. Rev. B \textbf{77}, 035305 (2008).

\bibitem{Studer09} M. Studer, S. Sch{\"o}n, K. Ensslin, and G.~Salis, Phys. Rev. B \textbf{79}, 045302 (2009).

\bibitem{Studer10} M.~Studer, M.\,P.~Walser, S.~Baer, H.~Rusterholz, S.~Sch{\"o}n, D.~Schuh, W.~Wegscheider, K.~Ensslin, and G.~Salis,
Phys. Rev. B 82, 235320 (2010).

\bibitem{IvchenkoGanichev} E.L. Ivchenko and S.D. Ganichev,
\textit{Spin Photogalvanics}
in \textit{Spin Physics in Semiconductors}, ed. M.I. Dyakonov (Springer 2008) pp. 245-277. 

\bibitem{GanichevPrettl} S.\,D.~Ganichev  and W.~Prettl, \textit{Intense Terahertz Excitation of Semiconductors} (Oxford
University Press, Oxford, 2006).




\bibitem{A1book} E.\,L.~Ivchenko  and  G.\,E.~Pikus, Superlattices and Other
Heterostructures. Symmetry and Optical Phenomena (Springer,
Berlin, 1997).


\bibitem{A1book2} E.\,L.~Ivchenko, Optical Spectroscopy of Semiconductor
Nanostructures (Alpha Science Int., Harrow, UK, 2005).


\bibitem{gyrotropy} 
We remind that the gyrotropic point group symmetry makes no difference between certain components 
of polar vectors, like electric current or electron momentum, and axial vectors, like a spin or magnetic field, 
and is described by the gyration tensor~\cite{Dyakonov08Ivchenko,Landau,Nye}. 
Gyrotropic  media are characterized by the linear in light or electron wavevector 
$\mathbf k$ spatial dispersion resulting in  optical activity (gyrotropy) or 
Rashba/Dresselhaus band spin-splitting in 
semiconductor structures ~\cite{Dyakonov08Ivchenko,Nye,AgranovichGinzburg,Kizel,physchem,cardona_review}, 
respectively. Among 21 crystal classes lacking inversion symmetry, 18 are gyrotropic, from which 
11 classes are enantiomorphic (chiral) and do not possess a reflection plane or rotation-reflection 
axis~\cite{Dyakonov08Ivchenko,Kizel,physchem}. Three nongyrotropic noncentrosymmetric classes 
are T$_d$, C$_{3h}$ and $D_{3h}$. We note that it is often, but misleading, stated that 
gyrotropy (optical activity) can be obtained only in non-centrosymmetric crystals having 
no mirror reflection plane. In fact 7 non-enantiomorphic classes groups 
(C$_{s}$, C$_{2v}$, C$_{3v}$, S$_{4}$, D$_{2d}$, C$_{4v}$ and C$_{6v}$) are gyrotropic 
allowing addressed above linear in $\bm k$ spin splitting, SGE, CPGE, MPGE excited by unpolarized radiation as well
as inversed SGE - spin orientation by the electric current, for review see e.g.~\cite{GanichevSchliemann}. 


\bibitem{Dyakonov08Ivchenko} E.L. Ivchenko and S.D.~Ganichev,
\textit{Spin photogalvanics}, in \textit{Spin Physics in Semiconductors}
ed. M.I.~Dyakonov,  (Springer, Berlin 2008), pp. 245-277.

\bibitem{Landau} L.D. Landau,  E.M. Lifshits and L.P. Pitaevskii, Vol.~8 \textit{Electrodynamics of Continuous Media} (Elsevier,
Amsterdam, 1984).

\bibitem{Nye} J.F. Nye \textit{Physical Properties of Crystals: Their Representation by Tensors and Matrices} (
Oxford Univ. Press, Oxford 1985). 

\bibitem{AgranovichGinzburg} V.M. Agranovich and V.L. Ginzburg, 
\textit{Crystal Optics with Spatial Dispersion, and Excitons} in Springer Series in Solid-State Sciences Vol. 42
(Springer, Berlin 1984). 

\bibitem{Kizel} V.A. Kizel', Yu.I. Krasilov, and V.I. Burkov, Usp. Fiz. Nauk \textbf{114}, 295 (1974) [Sov. Phys. Usp. \textbf{17}, 745 (1975)].

\bibitem{physchem} J. Jerphagnon and D.S. Chemla,  J. Chem. Phys. \textbf{65}, 1522 (1976).

\bibitem{cardona_review} B. Koopmans, P.V. Santos, and M. Cardona,
phys. stat. sol. (b) \textbf{205}, 419 (1998).

\bibitem{GanichevSchliemann} S.D. Ganichev, M. Trushin, and J. Schliemann, 
\textit{Spin orientation by electric current}, 
in \textit{Handbook of Spin Transport and Magnetism}
eds. E.Y. Tsymbal and I. Zutic  (Chapman and Hall 2011) pp. 487-497.

\bibitem{Sinitsin}
Fuxiang Li, Y.V. Pershin, V.A. Slipko, and N. A. Sinitsyn,
Phys. Rev. Lett. \textbf{111}, 067201 (2013).


\bibitem{Omega} Note that following to Ref.~\cite{omega1}
a $\bm k$ dependent Larmor precession frequency $\Omega (\bm{k})$ 
of electron spin precession around $B_{eff}(\bm{k})$ is 
commonly used in the literature~\cite{omega4,omega1,omega2,omega3,omega5}. 

\bibitem{omega1} G.E. Pikus and A.N. Titkov in {\it Optical Orientation}, edited by F.
Meier and B. P. Zakharchenya (North-Holland, Amsterdam, 1984).

\bibitem{omega2} S.V. Iordanskii, Yu.B. Lyanda-Geller, and G. E. Pikus, Pis'ma
Zh. Eksp. Teor. Fiz. {\bf 60}, 199 (1994) [JETP Lett. {\bf 60}, 206 (1994)].

\bibitem{omega3} F.G. Pikus and G.E. Pikus, Phys. Rev. B \textbf{51}, 16928 (1995).


\bibitem{omega5} L.E. Golub, Physics - Uspekhi {\bf 55}, 814 (2012)

\bibitem{A1Silva92} E.\,A.~de~Andrada~e~Silva, 
Phys. Rev.~B \textbf{ 46}, 1921 (1992).


\bibitem{Sandoval2013} M.\,A.~Toloza~Sandoval, A.~Ferreira~da~Silva,
E.\,A.~Andrada~e~Silva, and G.\,C.~La~Rocca, Phys. Rev. B
\textbf{87}, 081304(R) (2013).


\bibitem{combined} Note that due to combined effect of SIA and BIA in 
asymmetric (110) QWs an additional small term 
${H_{\rm SO} = \tilde{\alpha}(\sigma_x k_y + \sigma_y k_x)}$ is allowed. However, 
the constant $\tilde{\alpha}$ is about an order of magnitude smaller than the 
Rashba constant $\alpha$ in a rectangular QW subjected in electric 
field $\sim 10^5$~V/cm~\cite{Nestoklon_2012}. 
We also note that in symmetrically doped  (110)-grown QWs  there are spatial fluctuations of the 
Rashba constant yielding finite values of $\alpha$ which is zero in average. This 
spatially-fluctuating Rashba splitting leads to spin relaxation which limits the spin 
dephasing time in symmetrically-doped (110) QWs~\cite{Review_Glazov_Sherman_Dugaev}.


\bibitem{Nestoklon_2012} 
M. O. Nestoklon, S. A. Tarasenko, J.-M. Jancu,  P. Voisin,
hys. Rev. B
\textbf{85}, 205307 (2012).


\bibitem{Review_Glazov_Sherman_Dugaev} 
M.\,M.~Glazov, E.\,Ya.~Sherman, and V.\,K.~Dugaev, Physica E \textbf{42}, 2157 (2010).

\bibitem{Cartoixa2005_2} 
X.~Cartoixa, D.\,Z.\,Y.~Ting, Y.\,C.~Chang, Phys. Rev. B
\textbf{71}, 045313 (2005).

\bibitem{Vurgaftman2005} 
I. Vurgaftman and J. R. Meyer, J. Appl. Phys. \textbf{97}, 053707 (2005).

\bibitem{Sun2010} B. Y. Sun, P. Zhang, and M. W. Wu, J. Appl. Phys. \textbf{108}, 093709 (2010).




\bibitem{Ferreira91p9687}R.~Ferreira and G.~Bastard,
Phys. Rev. B. {\bf 43}, 9687 (1991).

\bibitem{Bastard93p439} G.~Bastard and R.~Ferreira, 
Europh. Lett. {\bf 23}, 439 (1993).

\bibitem{PRL2002bleching} S.D.~Ganichev, S.N.~Danilov, V.~V.~Bel'kov,
E.~L.~Ivchenko, M.~Bichler, W.~Wegscheider, D.~Weiss,
and W.~Prettl,
Phys. Rev. Lett. {\bf 88}, 057401-1 (2002).

\bibitem{Petra2004} P. Schneider, J. Kainz, S.D. Ganichev, V. V. Bel'kov, S. N. Danilov, M. M. Glazov, L. E. Golub, U. R{\"o}ssler, W. Wegscheider, D. Weiss, D. Schuh, and W. Prettl, 
J. Appl. Phys. \textbf{96}, 420 (2004). 

\bibitem{Korn2010} T.~Korn, M.~Kugler, M.~Griesbeck, R.~Schulz,
A.~Wagner, M.~Hirmer, C.~Gerl, D.~Schuh, W.~Wegscheider, and
C.~Sch{\"u}ller, New J. Phys. \textbf{12}, 043003 (2010).

\bibitem{Korn2010_2} T.~Korn, Phys. Reports \textbf{494}, 415 (2010).

\bibitem{Kugler2009} M.~Kugler, T.~Andlauer, T.~Korn, A.~Wagner,
S.~Fehringer, R.~Schulz, M.~Kubov\'{a}, C.~Gerl, D.~Schuh,
W.~Wegscheider, P.~Vogl, and C.~Sch{\"u}ller, Phys. Rev. B
\textbf{80}, 035325 (2009).


\bibitem{Koenig2007}
M. S. K{\"o}nig, S. Wiedmann, C. Brüne, A. Roth, H. Buhmann, L. Molenkamp, X. L. Qi, and S. C. Zhang, 
Science 318, 766 (2007).

\bibitem{Zhang} S. C. Zhang and X. L. Qi, 
Rev. Mod. Phys. 83, 1057 (2011). 


\bibitem{cubicGaN} Note that in bulk III--V semiconductors and \textit{cubic} GaN 
the constant $\beta$ is zero.


\bibitem{A1PRB02SiGe} S.\,D.~Ganichev, U.~R{\"o}ssler, W.~Prettl, E.\,L.~Ivchenko, V.\,V.~Bel'kov,
R.~Neumann, K.~Brunner,  and G.~Abstreiter, 
Phys.~Rev. B \textbf{ 66}, 075328 (2002).


\bibitem{A1Roessler02p313} U.~R{\"o}ssler and J.~Kainz, 
Solid State Commun. \textbf{121}, 313 (2002).

\bibitem{A1Wilamowski02p195315} Z.~Wilamowski, W.~Jantsch, H.~Malissa,  and
U.~R{\"o}ssler, 
Phys. Rev. B \textbf{ 66}, 195315 (2002).

\bibitem{A1Golub03bialike} L.\,E.~Golub and E.\,L.~Ivchenko,
Phys. Rev. B \textbf{ 69}, 115333 (2004).

\bibitem{Nestoklon2006} 
M. O. Nestoklon, L. E. Golub, and E. L. Ivchenko, \textbf{73}, 235334 (2006).

\bibitem{Nestoklon2008}  
M. O. Nestoklon, E. L. Ivchenko, J.-M. Jancu, and P. Voisin, 
Phys. Rev. B \textbf{77}, 155328 (2008).


\bibitem{Prada2011} M. Prada, G. Klimek, and R. Joynt, New J. Phys. \textbf{13}, 013009 (2011).





\bibitem{Belinicher} V.I. Belinicher, Phys. Lett. A {\bf 66}, 213
(1978).

\bibitem{sturman} B.I.~Sturman and V.M.~Fridkin, The Photovoltaic and
Photorefractive Effects in Non-Centrosymmetric Materials (Gordon
and Breach Science Publishers, New York, 1992).


\bibitem{Ivchenko89p175} E.L.~Ivchenko, Yu.B.~Lyanda-Geller, and
G.E.~Pikus,
Pis'ma Zh. Eksp. Teor. Fiz. \textbf{ 50}, 156  (1989)
[JETP Lett. \textbf{ 50}, 175 (1989)].

\bibitem{Nature02} S.\,D.~Ganichev, E.\,L.~Ivchenko,
V.\,V.~Bel'kov, S.\,A.~Tarasenko, M.~Sollinger, D.~Weiss,
W.~Wegscheider, and W.~Prettl, Nature (London) {\bf 417}, 153
(2002).

\bibitem{PRL01} S.\,D.~Ganichev, E.\,L.~Ivchenko, S.\,N.~Danilov, J.~Eroms,
W.~Wegscheider, D.~Weiss, and  W.~Prettl,
Phys. Rev. Lett. \textbf{ 86}, 4358 (2001).

\bibitem{Ganichev02} 
 S.\,D. Ganichev, E.\,L. Ivchenko, and W. Prettl,
Physica E {\bf 14}, 166 (2002).

\bibitem{BelkovJPCM}V.\,V.~Bel'kov, S.\,D.~Ganichev, E.\,L.~Ivchenko, S.\,A.~Tarasenko,
W.~Weber, S.~Giglberger, M.~Olteanu, H.-P.~Tranitz,
S.\,N.~Danilov, Petra~Schneider, W.~Wegscheider, D.~Weiss, and
W.~Prettl, J. Phys.: Condens. Matter \textbf{17}, 3405 (2005).

\bibitem{Golub2005} L.\,E.~Golub, Phys. Rev. B \textbf{71}, 235310 (2005).



\bibitem{Kohda2012} M. Kohda, V. Lechner, Y. Kunihashi, T. Dollinger, P. Olbrich, C. Sch{\"o}nhuber, I. Caspers, V.V. Bel'kov, L.E. Golub, D. Weiss, K. Richter, J. Nitta and S.D. Ganichev,
Phys. Rev. B \textbf{86}, 081306 (R) (2012).


\bibitem{inversedSGE} Inversed SGE, i.e. spin orientation by electric current (for recent review see~\cite{GanichevSchliemann}), 
can also be used for study of BIA/SIA anisotropy. This effect, given by $   S_l =  \sum_{m}Q^\prime_{l m} j_m $,
is observed in several low dimensional  systems~\cite{Ganichev04p0403641,Silov2004p5929,Kato2004p176601,Sih2005,Yang06p186605,Stern2006p126603} and 
like SGE reflects the spin splitting anisotropy, see e.g.~\cite{Aronov1991,Chaplik2002,Trushin07p155323,Raichev2007,Golub2011}.


\bibitem{Ganichev04p0403641}  S.D.~Ganichev, S.N.~Danilov, Petra~Schneider, V.V.~Bel'kov,
L.E.~Golub, W.~Wegscheider, D.~Weiss, and W.~Prettl,
cond-mat/0403641 (2004), see also
J. Magn. and Magn. Materials  \textbf{300}, 127 (2006).

\bibitem{Silov2004p5929} A.Yu. Silov, P.A. Blajnov,  J.H. Wolter,
R.~Hey, K.H.~Ploog and  N.S.~Averkiev, 
Appl. Phys. Lett. \textbf{ 85}, 5929 (2004).

\bibitem{Kato2004p176601} Y.K. Kato, R.C. Myers, A.C. Gossard and D.D.
Awschalom, 
Phys. Rev. Lett. \textbf{ 93}, 176601 (2004).

\bibitem{Sih2005} 
V.~Sih, R.\,C.~Myers, Y.\,K.~Kato, W.\,H.~Lau, A.\,C.~Gossard, and
D.\,D.~Awschalom, Nature Phys. \textbf{1},  31 (2005).


\bibitem{Stern2006p126603} N.P.~Stern, S.~Ghosh, G.~Xiang, M.~Zhu, N.~Samarth, and D.D.~Awschalom,
Phys. Rev. Lett. \textbf{97}, 126603 (2006).




\bibitem{Aronov1991} A.G. Aronov, Yu.B. Lyanda-Geller and G.E. Pikus, JETP \textbf{100}, 973
(1991) [Sov. Phys. JETP \textbf{73}, 537 (1991)].

\bibitem{Chaplik2002} A.V. Chaplik, M.V. Entin, and L.I. Magarill, Physica E \textbf{13}, 744
(2002).

\bibitem{Trushin07p155323} M. Trushin and J. Schliemann,
Phys. Rev. B \textbf{75}, 155323 (2007).

\bibitem{Raichev2007} O. E. Raichev, Phys. Rev. B \textbf{75}, 205340 (2007).

\bibitem{Golub2011} L.E. Golub and E.L. Ivchenko, Phys. Rev. B
 \textbf{84}, 115303 (2011).


\bibitem{belkovganichevreview} V.\,V.~Bel'kov, and S.\,D.~Ganichev, 
Semicond. Sci. and Tech. \textbf{23}, 114003 (2008).



\bibitem{BelkovGanichev} V.V. Bel'kov and S.D. Ganichev,
\textit{Zero-bias spin separation}, in \textit{Handbook of Spintronic Semiconductors}, 
eds. W.M.~Chen and I.A.~Buyanova (Pan Stanford Publishing 2010) pp. 243-265.

\bibitem{Winkler06p0605390} R. Winkler, \textit{Spin-Dependent Transport of Carriers
in Semiconductors}, in  \textit{Handbook of Magnetism and Advanced
Magnetic Materials} Vol. 5, Eds. H.~Kronmuller and S.~Parkin, 
(John Wiley \& Sons, NY 2007), arXiv:cond-mat/0605390.


\bibitem{Karch2010}  J. Karch, P. Olbrich, M.~Schmalzbauer, C. Zoth, C.~Brinsteiner, 
M. Fehrenbacher,  U.~Wurstbauer, M.\,M.~Glazov, S.A. Tarasenko, E.L. Ivchenko, D. Weiss, 
J.~Eroms, and S.D.~Ganichev,
Phys. Rev. Lett.  \textbf{97}, 227402 (2010).





\bibitem{Romashko2010} 
R. V. Romashko, A. I. Grachev, Yu. N. Kulchin, and A. A. Kamshilin, 
Optics Express, Vol. \textbf{18}, 27142 (2010).

\bibitem{Karch2011} J. Karch, C. Drexler, P. Olbrich, M. Fehrenbacher, M. Hirmer, M.M. Glazov, 
S.A. Tarasenko, E.L. Ivchenko, B. Birkner, J. Eroms, D. Weiss, R. Yakimova, S.Lara-Avila, S. Kubatkin, M. Ostler, T. Seyller, and S. D. Ganichev,
Phys. Rev. Lett. \textbf{107}, 276601 (2011).

\bibitem{Hosur2011} 
P. Hosur, Phys. Rev. B  \textbf{83}, 035309 (2011).

\bibitem{Kvon2011} Z.D. Kvon, S.N. Danilov, D.A. Kozlov, C. Zoth, N.N. Michailov, S.A. Dvoretzkii, and S.D. Ganichev,
JETP Lett. \textbf{94}, 816 (2011) [Pisma v ZhETP \textbf{94}, 895
(2011)].

\bibitem{McIverTI}  J. W. McIver, D. Hsieh, H. Steinberg, P. Jarillo-Herrero, and N. Gedik, 
Nature Nanotech. \textbf{7}, 96 (2012).

\bibitem{Dora_PGE_TI} 
B. Dora, J. Cayssol, F. Simon, and R. Moessner, Phys. Rev. Lett. \textbf{108}, 056602 (2012)


\bibitem{Wu2012} 
Q. S. Wu, S. N. Zhang, Z. Fang, X. Dai, Physica E \textbf{44},   895 (2012).


\bibitem{Drexler2013} C.~Drexler,  S.~A.~Tarasenko, P.~Olbrich, J.~Karch,
M.~Hirmer, F. M\"{u}ller, M.~Gmitra, J. Fabian,
R.~Yakimova, S.~Lara-Avila, S.~Kubatkin, and S.~D.~Ganichev,
Nature Nanotechnology \textbf{8}, 104 (2013).


\bibitem{PRB_HgTe}  P.\,Olbrich, C.\,Zoth, P.\,Vierling, K.-M.\,Dantscher,
G.V.\,Budkin, S.A.\,Tarasenko, V.V.\,Bel'kov, D.A.\,Kozlov,  Z.D.\,Kvon,
N.N.\,Mikhailov, S.A.\,Dvoretsky, and S.D.\,Ganichev,
Phys. Rev. B  \textbf{87}, 235439 (2013).

\bibitem{Artemenko2013} 
S. N. Artemenko and V. O. Kaladzhyan, JETP Lett. \textbf{97}, 82 (2013).


















\bibitem{Lusakowski2003} A.~{\L}usakowski, J.~Wr{\'o}bel, and T.~Dietl,
Phys. Rev. B {\bf 68}, R081201 (2003).



\bibitem{Kettemann2007} S.~Kettemann, Phys. Rev. Lett. \textbf{98}, 176808
(2007).

\bibitem{Flatte}W.\,H.~Lau and  M.\,E.~Flatt\'{e}, Phys. Rev. B \textbf{72}, R161311 (2005).



\bibitem{Ganichev95DX}  S.~D.~Ganichev, J.~Diener,  I.~N.~Yassievich, W.~Prettl,
B.~K.~Meyer, and K.~W.~Benz,
Phys. Rev. Lett. {\bf  75},  1590 (1995).

\bibitem{Ziemann2000}  E.~Ziemann, S.~D.~Ganichev, I.~N.~Yassievich, V.~I.~Perel, and
W.~Prettl,
J. Appl. Phys. {\bf 87}, 3843 (2000).

\bibitem{GanichevPrettl2002} S.D.~Ganichev, I.~N.~Yassievich, and  W.~Prettl,
J. Phys.: Condens. Matter {\bf 14}, R1263 (2002).

\bibitem{Kvon2008} Z.D. Kvon, S.N. Danilov, N.N. Mikhailov, S.A. Dvoretsky, and S.D.Ganichev, 
Physica E \textbf{40}, 1885 (2008).


\bibitem{Knippels99p1578}  G.\,M.\,H. Knippels, X.~Yan, A.\,M.~MacLeod,
W.\,A.~Gillespie, M.~Yasumoto, D.~Oepts, and A.\,F.\,G. van
der~Meer,
Phys. Rev. Lett. \textbf{ 83}, 1578 (1999).

\bibitem{Svelto2010}  O.~Svelto, Principles of Lasers (Springer, 5th ed., 2010).

\bibitem{PRB2008}  W.~Weber, L.E. Golub, S.N. Danilov, J. Karch, C. Reitmaier, B. Wittmann, V.V. Bel'kov, E.L.~Ivchenko,  
Z.D. Kvon, N.Q. Vinh, A.F.G. van der Meer, B. Murdin, and S.D.~Ganichev, Phys. Rev. B \textbf{77}, 245304 (2008). 

\bibitem{Wittmann2008} B. Wittmann, L. E. Golub, S. N. Danilov, J. Karch, C. Reitmaier, Z. D. Kvon, N. Q. Vinh, A. F. G. van der Meer, B. Murdin, S. D. Ganichev, 
Phys. Rev. B \textbf{78}, 205435 (2008).

\bibitem{Wittmann2010} B. Wittmann, S. N. Danilov, V. V. Bel'kov, S. A. Tarasenko, E. G. Novik, H. Buhmann, C. Br{\"u}ne, L. W. Molenkamp, E. L. Ivchenko, Z. D. Kvon, N. N. Mikhailov, S. A. Dvoretsky, N. Q. Vinh, A.\,F.\,G.~van~der~Meer, B. Murdin, and S. D. Ganichev, 
Sci. and Technology \textbf{25}, 095005 (2010).

\bibitem{Danilov2009} S. N. Danilov, B. Wittmann, P. Olbrich, W. Eder, W. Prettl, L. E. Golub, E. V. Beregulin, Z. D. Kvon, N. N. Mikhailov, S. A. Dvoretsky, V. A. Shalygin, N. Q. Vinh, A. F. G. van der Meer, B. Murdin, and S. D. Ganichev, 
J. Appl. Phys. \textbf{105}, 013106 (2009).

\bibitem{Jiang2011_1}  Ch.~Jiang, H.~Ma, J.~Yu, Y.~Liu, and Y.~Chen,
Appl. Phys. Lett. \textbf{99}, 032106 (2011).

\bibitem{Belkov2003} V.V.~Bel'kov, S.D.~Ganichev, Petra~Schneider, C.~Back,
M.~Oestreich, J.~Rudolph, D.~H{\"a}gele, L.E.~Golub,
W.~Wegscheider, W.~Prettl, 
Solid State Communic. {\bf 128}, 283 (2003).


\bibitem{Cho2007_1} K. S. Cho, Y. F. Chen, Y. Q. Tang, and B. Shen, 
Appl. Phys. Lett. \textbf{90}, 041909 (2007).  

\bibitem{Dai2010} J.~Dai, H.-Z.~Lu, C.L. Yang,
S.-Q.~Shen, F.-C.~Zhang, and X.~Cui, Phys. Rev. Lett. \textbf{104}, 246601 (2010).



\bibitem{Zhang2010ZnO}  Q.~Zhang, X.\,Q.~Wang, C.\,M.~Yin, F.\,Y.~Xu,
N.~Tang, B.~Shen, Y.\,H.~Shen, K.~Chang, W.\,K.~Ge, Y.~Ishitani,
and A.~Yoshikawa,  Appl. Phys. Lett. \textbf{97}, 041907 (2010).


\bibitem{Priyadarshi2013} S.~Priyadarshi, K.~Pierz, and M.~Bieler,
Appl. Phys. Lett. \textbf{102}, 112102 (2013).


\bibitem{Sakai} Kiyomi Sakai \textit{Terahertz Optoelectronics} (Topics in Applied Physics, Springer 2005).

\bibitem{Lee} Yun-Shik Lee,  \textit{Principles of Terahertz Science and Technology}  (Springer 2009).

\bibitem{Drexler2010}  C. Drexler, V.V. Bel'kov, B. Ashkinadze, P. Olbrich, C. Zoth,
V. Lechner, Ya.V. Terent'ev, D.R.~Yakovlev,
G. Karczewski, T. Wojtowicz, D. Schuh,  W. Wegscheider,  and S.~D.~Ganichev,
Appl. Phys. Lett. \textbf{97}, 182107 (2010).


\bibitem{zerobias} S.D. Ganichev, V.V. Bel'kov, S.A. Tarasenko, S.N. Danilov, S. Giglberger, Ch. Hoffmann, E.L. Ivchenko, 
D. Weiss, W. Wegscheider, Ch. Gerl, D. Schuh, J. Stahl, J. De Boeck, G.~Borghs, and W. Prettl,
Nature Physics  \textbf{2}, 609 (2006). 


\bibitem{sun2010}
D.~Sun, C.~Divin, J.~Rioux, J.~E. Sipe, C.~Berger, W.~A. de~Heer, P.~N. First,
  T.~B. Norris.
Nano Lett. \textbf{10}, 1293 (2010).

\bibitem{Sun:2012ys}
D.~Sun, J.~Rioux, J.~E. Sipe, Y.~Zou, M.~T. Mihnev, C.~Berger, W.~A. de~Heer,
  P.~N. First, and T.~B. Norris.
Phys. Rev. B {\bf 85}, 165427 (2012).

\bibitem{Auston1} P.R. Smith, D.H. Auston, and M.C. Nuss, 
IEEE J. Quant. Electron. {\bf QE-24}, 255 (1988).

\bibitem{Auston2} X.-C. Zhang, B.B. Hu, J.T. Darrow, and D.H. Auston, 
Appl. Phys. Lett. {\bf 56}, 1011 (1990).


\bibitem{shapeQW} Note that the influence of QW shape on photogalvanic 
and magneto-photogalvanic effects
have been analyzed in~\cite{Majchrowski2009,Majchrowski2009_2,Entin2013}.

\bibitem{Majchrowski2009} 
K. Majchrowski, W. Pasko, and I. Tralle, Phys. Lett. A \textbf{373} 2959 (2009).
\bibitem{Majchrowski2009_2} 
K. Majchrowski, W. Pasko, and I. Tralle, Acta Phys. Polon. A \textbf{116}, 854 (2009).
\bibitem{Entin2013} 
M. V. Entin and L. I. Magarill, JETP Lett. \textbf{97}, 639 (2013).






\bibitem{Walzer2012} M. P. Walser, U. Siegenthaler, V. Lechner, D. Schuh, S. D. Ganichev, W. Wegscheider, and G. Salis,
Phys. Rev. B \textbf{86}, 195309 (2012).

\bibitem{Walzer2012_2}  P. Walser, C. Reichl, W. Wegscheider, and G. Salis, Nat. Phys.
8, 757 (2012).


\bibitem{Snelling91} M. J. Snelling, G.~P.~Flinn, A.~S.~Plaut, R.~T. Harley, A.~C.~Tropper, R.~Eccleston,
and C.~C.~Phillips,
Phys. Rev. B \textbf{44}, 11345 (1991).

\bibitem{Ivchenko_g_factor}  E.~L.~Ivchenko, A. A. Kiselev, and M. Willander,
Sol. St. Com. \textbf{102}, 375 (1997).

\bibitem{SalisNat01} G. Salis, Y. Kato, K. Ensslin, D.~C. Driscoll,
A.~C. Gossard, and D. ~D. Awschalom, Nature \textbf{414}, 619
(2001).

\bibitem{yugova} I. A. Yugova, A. Greilich, D.~R. Yakovlev, A.~A. Kiselev, M. Bayer,
V.~V. Petrov, Yu.~K. Dolgikh, D. Reuter, and A.~D. Wieck,
Phys. Rev. B \textbf{75},
245302 (2007).

\bibitem{Kuglerprb09} M. Kugler, T. Andlauer, T. Korn, A. Wagner, S. Fehringer, R. Schulz, M. Kubova,
C. Gerl, D. Schuh, W. Wegscheider, P. Vogl, and C.~Sch\"{u}ller,
Phys. Rev. B \textbf{80},  035325 (2009).  

\bibitem{Lechner2011} V. Lechner, L.E. Golub, F. Lomakina, V.V. Bel'kov, P. Olbrich, S. Stachel, I. Caspers, M. Griesbeck, M. Kugler, M.J. Hirmer, T. Korn, C. Sch{\"u}ller, D. Schuh, W. Wegscheider, and S.D. Ganichev, 
Phys. Rev. B \textbf{83}, 155313 (2011).


\bibitem{Tarasenko_orbital} S.~A.~Tarasenko,
Phys. Rev. B \textbf{77}, 085328 (2008). 

\bibitem{Lommer88p728} G.~Lommer, F.~Malcher, and U.~R{\"o}ssler,
Phys. Rev. Lett. {\bf 60}, 728 (1988).

\bibitem{Nitta97p1335} J.~Nitta, T.~Akazaki, and  H.~Takayanagi,
Phys. Rev. Lett. \textbf{ 78}, 1335 (1997).



\bibitem{Li2012} 
Juerong Li, A.\,M.~Gilbertson, K.\,L.~Litvinenko, L.\,F.~Cohen,
and S.\,K.~Clowes, Phys. Rev. B \textbf{85}, 045431 (2012).

\bibitem{Stachel2012} S. Stachel, P. Olbrich, C. Zoth, U. Hagner, 
T. Stangl, C. Karl, P. Lutz, V.V. Bel'kov, S.K. Clowes, T. Ashley, 
A.M. Gilbertson, and S.D. Ganichev,
Phys. Rev. B \textbf{85}, 045305 (2012).

\bibitem{Diehl2007MPGE001} H. Diehl, V. A. Shalygin, S. N. Danilov, S. A. Tarasenko, V. V. Bel'kov, D. Schuh, W. Wegscheider, W. Prettl and S. D. Ganichev, 
J. Phys.: Condens. Matter \textbf{19}, 436232 (2007). 

\bibitem{Diehl2009} H. Diehl, V. A. Shalygin, L. E. Golub, S. A. Tarasenko, S. N. Danilov, V. V. Bel'kov, E. G. Novik, H. Buhmann, L. W. Molenkamp, C. Br{\"u}ne, E. L. Ivchenko, and S. D. Ganichev, 
Phys. Rev. B, \textbf{80}, 075311 (2009).

\bibitem{Khodaparast2004} 
G.\,A.~ Khodaparast, R.\,E.~Doezema, S.\,J.~Chung,
K.\,J.~Goldammer, and M.\,B.~Santos Phys. Rev. B \textbf{70},
155322 (2004).

\bibitem{Gilbertson2009} 
A.\,M.~Gilbertson, W.\,R.~Branford, M.~Fearn, L.~Buckle,
P.\,D.~Buckle, T.~Ashley, and L.\,F.~Cohen, Phys. Rev. B
\textbf{79}, 235333 (2009).

\bibitem{Kallaher2010} R.\,L.~Kallaher, J.\,J.~Heremans, N.~Goel,
S.\,J.~Chung, and M.\,B.~Santos, Phys. Rev B \textbf{81}, 075303
(2010).

\bibitem{Akabori2008} 
M.~Akabori1, V.\,A.~Guzenko, T.~Sato, Th.~Sch{\"a}pers, T.~Suzuki,
and S.~Yamada, Phys. Rev. B \textbf{77}, 205320 (2008).

\bibitem{Leontiadou2011} 
M.\,A.~Leontiadou, K.\,L.~Litvinenko, A.\,M.~Gilbertson, C.\,R.~
Pidgeon, W.\,R.~Branford, L.\,F.~Cohen, M.~Fearn, T.~Ashley,
M.\,T.~Emeny, B.\,N.~Murdin, and S.\,K.~Clowes, J. Phys.: Condens.
Matter \textbf{23}, 035801 (2011).





\bibitem{Golub_2003} L. E. Golub, Physica E \textbf{17}, 342 (2003).

\bibitem{Duc2010} H. T. Duc, J. F{\"o}rstner,and T. Meier,
Phys. Rev. B \textbf{82}, 115316 (2010).

\bibitem{Lu2011} H.-Z.~Lu, B.~Zhou, F.-C.~Zhang, and S.-Q.~Shen,
Phys. Rev. B \textbf{83}, 125320 (2011).

\bibitem{Yu2011} J.\,L~Yu, Y.\,H.~Chen, C.\,Y.~Jiang, Y.~Liu, and H.~Ma,
J. Appl. Phys. \textbf{109}, 053519 (2011).

\bibitem{Yu2012_2} J.\,L.~Yu, Y.\,H.~Chen, C.\,Y.~Jiang, Y.~Liu,
H.~Ma, and L.\,P.~Zhu, Appl. Phys. Lett. \textbf{100}, 142109 (2012).

\bibitem{Yu2013} 
J. Yu, Y. Chen, S. Cheng, and Y. Lai, Physica E \textbf{49}, 92 (2013).


\bibitem{Tarasenko02p552} S.\,A.~Tarasenko  and N.\,S.~Averkiev,
Pis'ma ZhETF {\bf 75}, 669 (2002) [ JETP Lett. {\bf 75}, 552
(2002)].

\bibitem{Averkiev2005} N.\,S.~Averkiev, M.\,M.~Glazov, S.\,A.~Tarasenko,
Sol. State Commun. \textbf{133} 543 (2005).

\bibitem{Desrat2006} W.~Desrat, D.\,K.~Maude, Z.\,R.~Wasilewski, R.~Airey,
and G.~Hill, Phys. Rev. B \textbf{74}, 193317 (2006).



\bibitem{Liu2006} 
M.-H.~Liu, K.-W.~Chen, S.-H.~Chen, and C.-R.~Chang, Phys. Rev. B
\textbf{74}, 235322 (2006).

\bibitem{Cheng2006} J.\,L.~Cheng and M.\,W.~Wu, J. Appl. Phys. \textbf{99},
083704 (2006).

\bibitem{Bernevig2008} 
B. A. Bernevig and J. Hu, Phys. Rev. B \textbf{78}, 245123 (2008).

\bibitem{Li2010} J.~Li and K.~Chang, Phys. Rev. B \textbf{82}, 033304 (2010).

\bibitem{Slipko}
V.A. Slipko, I. Savran, Yu. V. Pershin, 
Phys. Rev. B \textbf{83}, 193302 (2011).

\bibitem{Slipko2013} 
V. A. Slipko, A. A. Hayeva, and Yu. V. Pershin, Phys. Rev. B \textbf{87}, 035430 (2013).








\bibitem{doping_dress} Note that the electron density and sample 
temperature can also affect both linear and cubic 
in $\bm k$ Dresselhaus terms, see Eq.~\eqref{Beff3},\eqref{Beff1n} and 
related discussion.  While this influence is small in materials with 
weak cubic in $\bm k$ spin splitting in narrow band semiconductors 
like InAs-based QWs it may play an important role, see section~\ref{helixexp}.


\bibitem{Engels97}  G. Engels, J. Lange, Th. Sch\"apers, and H. L\"uth, Phys. Rev. B {\bf 55}, R1958 (1997).


\bibitem{Koga02}    T. Koga, J. Nitta, T. Akazaki, and H. Takayanagi, Phys. Rev. Lett. {\bf 89}, 046801 (2002).

\bibitem{FanielKoga2011} S. Faniel, T. Matsuura, S. Mineshige, Y. Sekine, and T. Koga,
Phys. Rev. B \textbf{83}, 115309 (2011).

\bibitem{Glazov06} M. M. Glazov and L. E. Golub, Semicond. {\bf 40}, 1209 (2006).

\bibitem{Stanescu07}  T.D. Stanescu and V. Galitski, Phys.~Rev. B {\bf 75}, 125307 (2007).

\bibitem{Duckheim10} M. Duckheim, D. Loss, M. Scheid, K. Richter, I.
       Adagideli, and P. Jacquod, Phys. Rev. B {\bf 81}, 085303 (2010).

\bibitem{Luffe11} M.C. L\"uffe, J. Kailasvuori, and T.S. Nunner,  Phys. Rev. B {\bf 84}, 075326 (2011).



                        

\bibitem{Faniel11} S. Faniel, T. Matsuura, S. Mineshige, Y. Sekine, and T. Koga,
Phys. Rev. B {\bf 83}, 115309 (2011).


\bibitem{Sch08} M. Scheid, M. Kohda, Y. Kunihashi, K. Richter, and J. Nitta,
Phys. Rev. Lett. {\bf 101}, 266401 (2008).



\bibitem{Hassenkam1997} T.~Hassenkam, S.~Pedersen, K.~Baklanov, A.~Kristensen,
C.\,B.~Sorensen, P.\,E.~Lindelof, F.\,G.~Pikus, G.\,E.~Pikus,
Phys. Rev. B \textbf{55}, 9298 (1997). 

\bibitem{Hall2003PRB} 
K.\,C.~Hall, K.~G{\"u}ndo\v{g}du, E.~Altunkaya, W.\,H.~Lau,
M.\,E.~Flatt\'{e}, T.\,F.~Boggess, J.\,J.~Zinck,
W.\,B.~Barvosa-Carter, and S.\,L.~Skeith, Phys. Rev. B
\textbf{68}, 115311 (2003).

\bibitem{Henini2004} 
M.~Henini, O.\,Z.~Karimov, G.\,H.~John, R.\,T.~Harley, and
R.\,J.~Airey, Physica E \textbf{23} 309 (2004).

\bibitem{Morita2005} 
K.~Morita, H.~Sanada, S.~Matsuzaka, C.\,Y.~Hu, Y.~Ohno, and
H.~Ohno, Appl. Phys. Lett. \textbf{87}, 171905 (2005).


\bibitem{Ku2005} 
K.\,C.~Ku, S.\,H.~Chun, W.\,H.~Wang, W.~Fadgen, D.\,A.~Issadore,
N.~Samarth, R.\,J.~Epstein, and D.\,D.~Awschalom, J. of
Superconduct. \textbf{18}, 185 (2005).

\bibitem{Hicks2006} 
J.~Hicks, K.~G{\"u}ndo\v{g}du, A.\,N.~Kocbay, K.\,C.~Hall,
T.\,F.~Boggess, K.~Holabird, A.~Hunter, and J.\,J.~Zinck, Physica
E \textbf{34}, 371 (2006).

\bibitem{Couto2007} 
O.\,D.\,D.~Couto, Jr., F.~Iikawa, J.~Rudolph, R.~Hey, and
P.\,V.~Santos, Phys. Rev. Lett. \textbf{98}, 036603 (2007).

\bibitem{Schreiber2007} 
L.~Schreiber, D.~Duda, B.~Beschoten, G.~G{\"u}ntherodt,
H.-P.~Sch{\"o}nherr, and J.~Herfort, phys. stat. solidi (b)
\textbf{244}, 2960 (2007).


\bibitem{Schreiber2007_2} 
L.~Schreiber, D.~Duda, B.~Beschoten, G.~G{\"u}ntherodt,
H.-P.~Sch{\"o}nherr, and J. Herfort, Phys. Rev. B \textbf{75},
193304 (2007).

\bibitem{Eldridge2010_4} 
S.~Eldridge, P.\,G.~Lagoudakis, M.~Henini, and R.\,T.~Harley,
Phys. Rev. B \textbf{81}, 033302 (2010).

\bibitem{Iba2010} 
S.~Iba, S.~Koh, and H.~Kawaguchi, Appl. Phys. Lett. \textbf{97}
202102 (2010).

\bibitem{Voelkl2011} 
R.~V{\"o}lkl, M.~Griesbeck, S.\,A.~Tarasenko, D.~Schuh,
W.~Wegscheider, C.~Sch{\"u}ller, and T.~Korn, Phys. Rev. B
\textbf{83}, 241306(R) (2011).

\bibitem{Huebner2011} J.~H{\"u}bner, S.~Kunz, S.~Oertel, D.~Schuh, M.~Pochwa{\l}a,
H.\,T.~Duc, J.~F{\"o}rstner, T.~Meier, and M.~Oestreich, Phys.
Rev. B \textbf{84}, 041301(R) (2011). 





\bibitem{Wu2002} 
M.\,W.~Wu, M.~Kuwata-Gonokami, Sol. State Commun. \textbf{121},
509 (2002).


\bibitem{Chang2005} 
S.\,W.~Chang and S.\,L.~Chuang, Phys. Rev B \textbf{72} 115429
(2005).

\bibitem{Tarasenko2009PRB80} S.\,A.~Tarasenko,  Phys. Rev. B \textbf{80}, 165317 (2009).


\bibitem{Glazov2010} 
M.\,M.~Glazov, M.\.A.~Semina, and E.\,Ya.~Sherman, Phys. Rev. B
\textbf{81}, 115332 (2010).

\bibitem{Zhou2010} 
Y.~Zhou and M.\,W.~Wu, EPL \textbf{89}, 57001 (2010).

\bibitem{Poshakinskiy2013} A.\,V.~Poshakinskiy and S.\,A.~Tarasenko,
Phys. Rev. B \textbf{87}, 235301 (2013). 


\bibitem{Olbrich2009_110} P. Olbrich, J. Allerdings, V. V. Bel'kov, S. A. Tarasenko, D. Schuh, W. Wegscheider, T. Korn, C. Sch{\"u}ller, D. Weiss, and S. D. Ganichev, 
Phys. Rev. B 79, 245329 (2009). 

\bibitem{Zhao2005} 
H.~Zhao, X.~Pan, A.\,L.~Smirl, R.\,D.\,R.~Bhat, A.~Najmaie,
J.\,E.~Sipe, and H.\,M.~van~Driel, Phys. Rev. B \textbf{72},
201302(R) (2005).

\bibitem{Shalygin2006} V. A. Shalygin, H. Diehl, Ch. Hoffmann, S. N. Danilov, T. Herrle, S. A. Tarasenko, D. Schuh, Ch. Gerl, W. Wegscheider, W. Prettl and S. D. Ganichev, 
JETP Lett. \textbf{84}, 570 (2006). 

\bibitem{Diehl2007_110} H. Diehl, V. A. Shalygin, V. V. Bel'kov, Ch. Hoffmann, S. N. Danilov, T. Herrle, S. A. Tarasenko, D. Schuh, Ch. Gerl, W. Wegscheider, W. Prettl, and S. D. Ganichev, 
New J. Physics \textbf{9}, special issue \textit{Focus on Spintronics in Reduced Dimensions}, 349 (2007). 



\bibitem{Hu2008} 
K.\,G.~Hu, Sol. State Commun. \textbf{148}, 283 (2008).


\bibitem{Nakamurabook} S. Nakamura, S. Pearton, and G. Fasol, 
\textit{The Blue Laser Diode. The Complete Story}  
(Springer, Berlin, 2007).

\bibitem{GaN1} Q. Chen, M. Asif Khan, J. W. Yang, C. J. Sun, M. S. Shur, and H. Park,
Appl. Phys. Lett. \textbf{69}, 794 (1996).

\bibitem{GaN2}  Y.-F. Wu, B. P. Keller, S. Keller, D. Kapolnek, P. Kozodoy, S. P. Denbaars,
and U. K. Mishra, Appl. Phys. Lett. \textbf{69}, 1438 (1996).
\bibitem{GaN3}  O. Ambacher, J. Smart, J. R. Shealy, N. G. Weimann, K. Chu, M. Murphy,
W. J. Schaff, L. F. Eastman, R. Dimitrov, L. Wittmer, M. Stutzmann, W.
Rieger, and J. Hilsenbeck, J. Appl. Phys. \textbf{85}, 3222 (1999). 

\bibitem{Dietl2000} T. Dietl, H. Ohno, F. Matsukura, J. Cibert, and D. Ferrand, Science \textbf{287}, 1019 (2000).  

\bibitem{7gado} S. Dhar, O. Brandt, M. Ramsteiner, V. F. Sapega, and K. H. Ploog, Phys.
Rev. Lett. \textbf{94}, 037205 (2005).

\bibitem{7gadobis} 
G.M.~Dalpian, Su-Huai-Wei, Phys. Rev. B \textbf{72}, 115201 (2005).

\bibitem{8gado} S. Dhar, T. Kammermeier, A. Ney, L. Pérez, K. H. Ploog, A. Melnikov,
and A. D. Wieck, Appl. Phys. Lett. \textbf{89}, 062503 (2006).

\bibitem{9gado} M. A. Khaderbad, S. Dhar, L. Pérez, K. H. Ploog, A. Melnikov, and A. D.
Wieck, Appl. Phys. Lett. \textbf{91}, 072514 (2007). 

\bibitem{Buss2013Gd}
 J. H. Buss, J. Rudolph, S. Shvarkov, F. Semond, 
 D. Reuter, A.D. Wieck, D. H{\"a}gele, Appl. Phys. Lett. 103, 92401 (2013).


\bibitem{Beschoten01}
B. Beschoten, E. Johnston-Halperin, D. K. Young, M. Poggio, J. E. Grimaldi, 
S. Keller, S. P. DenBaars, U. K. Mishra,  E. L. Hu,  
and D. D. Awschalom, Phys. Rev. B \textbf{63}, 121202 (2001). 

\bibitem{Buss2010} J. H. Bu\ss, J. Rudolph, F. Natali, F. Semond, and D. H{\"a}gele, Phys. Rev. B \textbf{81}, 155216 (2010).

\bibitem{Buss2011} J. H. Bu\ss, J. Rudolph, S. Starosilec, A. Schaefer, F. Semond, Y. Cordier, A. D. Wieck, and D. H{\"a}gele, Phys. Rev. B \textbf{84}, 153202 (2011).

\bibitem{WeberAPL2005GaN} W. Weber, S. D. Ganichev, Z. D. Kvon, V. V. Bel'kov, L. E. Golub, S. N. Danilov, 
D. Weiss, W. Prettl, H.-I. Cho, and J.-H. Lee, 
Appl. Phys. Lett. \textbf{87}, 262106 (2005). 

\bibitem{CingolaniPRB2000} R. Cingolani, A.~Botchkarev, H.~Tang, H.~Morkoc,
G.~Traetta, G.~Coli, M.~Lomascolo, A.~Di~Carlo, F.~Della~Sala, and
P. ~Lugli,
Phys. Rev. B \textbf{61}, 2711 (2000).

\bibitem{Litvinov}  V.\,I. Litvinov, Phys. Rev. B \textbf{68}, 155314 (2003).



\bibitem{HeAPL2007} X.\,W.~He, B.~Shen, Y.\,Q.~Tang, N.~Tang, C.\,M.~Yin,
F.\,J.~Xu, Z.\,J.~Yang, G.\,Y.~Zhang, Y.\,H.~Chen, C.\,G.~Tang,
and Z.\,G.~Wang,
Appl. Phys. Lett. \textbf{91}, 071912 (2007).

\bibitem{TangApl2007} Y.\,Q.~Tang, B.~Shen, X.\,W.~He, K.~Han,
N.~Tang, W.\,H.~Chen, Z.\,J.~Yang, G.\,Y.~Zhang, Y.\,H.~Chen,
C.\,G.~Tang, Z.\,G.~Wang, K.\,S.~Cho, and Y.\,F.~Chen,
Appl. Phys. Lett. \textbf{91}, 071920 (2007).

\bibitem{ChoPRB2007} K.\,S.~Cho, C.-T.~Liang, Y.\,F.~Chen,
Y.\,Q.~Tang, and B.~Shen,
Phys. Rev. B \textbf{75}, 071912 (2007).



\bibitem{SSC07} W.~Weber, S.~Seidl,  V.\,V. Bel'kov, L.\,E.~Golub,
E.\,L.~Ivchenko, W.~Prettl, Z.\,D.~Kvon, Hyun-Ick~Cho,
Jung-Hee~Lee, and S.\,D.~Ganichev,
Solid State Comm. \textbf{145}, 56 (2008).



\bibitem{Tang2008} N. Tang, B. Shen, K. Han, F.-C. Lu, F.-j. Xu, Z.-X. Qin, and G.-Y. Zhang, Appl. Phys. Lett. \textbf{93}, 172113 (2008).







\bibitem{ChoApl2005} K.\,S. Cho, Tsai-Yu~Huang, Hong-Syuan~Wang, Ming-Gu~Lin, Tse-Ming~Chen,
C.-T.~Liang, Y.\,F. ~Chen, and Ikai~Lo,
Appl. Phys. Lett. \textbf{86}, 222102 (2005).


\bibitem{ThillosenAPL2006} N. Thillosen, Th.~Sch{\"a}pers,
N.~Kaluza, H.~Hardtdegen, and V.\,A.~Guzenko,
Appl. Phys. Lett. \textbf{88}, 022111 (2006).

\bibitem{SchmultPRB2006} S. Schmult, M.\,J.~Manfra,
A.~Punnoose, A.\,M.~Sergent, K.\,W.~Baldwin, and R.\,J.~Molnar,
Phys. Rev. B \textbf{74}, 033302 (2006).

\bibitem{TangAPL2006} N.~Tang, B.~Shen, M.\,J.~Wang, K.~Han, Z.\,J.~Yang,
K.~Xu, G.\,Y.~Zhang, T.~Lin, B.~Zhu, W.\,Z.~Zhou, and J.\,H.~Chu,
Appl. Phys. Lett. \textbf{88}, 172112 (2006).

\bibitem{Spirito2011} D. Spirito, L. Di~Gaspare, G. Frucci, F. Evangelisti, A. Di~Gaspare, A. natargiacomo, E. Giovine, S. Roddaro, and F. Beltram, Phys. Rev. B \textbf{83}, 155318 (2011).

\bibitem{Zhang2008InN} Z. Zhang, R. Zhang, B. Liu, Z. L. Xie, X. Q. Xiu,
P. Han, H. Lu, Y. D. Zheng, Y. H. Chen, C. G. Tang, and Z. G. Wang,
Sol. State Commun. \textbf{145}, 159 (2008).

\bibitem{Zhang2009InN_1} 
Q. Zhang, X. Q. Wang, X. W. He, C. M. Yin, F. J. Xu, B. Shen, Y. H. Chen, Z. G. Wang, Y. Ishitani, and A. Yoshikawa, Appl. Phys. Lett. \textbf{95}, 031902 (2009).

\bibitem{Zhang2009InN} Z. Zhang, R. Zhang, Z. L. Xie, B. Liu, M. Li,
D. Y. Fu, H. N. Fang, X. Q. Xiu, H. Lu, Y. D. Zheng, Y. H. Chen,
C. G. Tang, and Z. G. Wang, Sol. State Commun. \textbf{149}, 1004 (2009).


\bibitem{Duan2013_ZnO} 
J. X. Duan, N. Tang, J. D. Ye, F. H. Mei, K. L. Teo, Y. H. Chen, W. K. Ge, and B. Shen, Appl. Phys. Lett. \textbf{102}, 192405 (2013).























\bibitem{Malissa2004} 
H. Malissa, W. Jantsch, M. M{\"u}hlberger, F. Sch{\"a}ffler, Z. Wilamowski, M. Draxler, and P. Bauer, Appl. Phys. Lett. \textbf{85}, 1739 (2004).

\bibitem{Tyryshkin2005} 
A. M. Tyryshkin, S. A. Lyon, W. Jantsch, and F. Sch{\"a}ffler, Phys. Rev. Lett. \textbf{94}, 126802 (2005).

\bibitem{Matsunami2006} 
J. Matsunami, M. Ooya, and T. Okamoto, Phys. Rev. Lett. \textbf{97}, 066602 (2006).

\bibitem{PureSpin2007} 
S. D. Ganichev, S. N. Danilov, V. V. Bel'kov, S. Giglberger, S. A. Tarasenko, E. L. Ivchenko, 
D. Weiss, W. Jantsch, F. Sch{\"a}ffler, D. Gruber, and W. Prettl, Phys. Rev. B \textbf{75}, 155317 (2007) 

\bibitem{Wei2007} C. M. Wei. K. S. Cho, Y. F. Chen, Y. H. Peng,
C. W. Chiu, and C. H. Kuan, Appl. Phys. Lett. \textbf{91}, 252102 (2007).

\bibitem{Detector2007} S. D. Ganichev, J. Kiermaier, W. Weber, S. N. Danilov, D. Schuh, 
Ch. Gerl, W. Wegscheider, D. Bougeard, G. Abstreiter, and W. Prettl, 
Appl. Phys. Lett. \textbf{91}, 091101 (2007). 











\bibitem{GanichevAPL2000}S.~D.~Ganichev, E.~L.~Ivchenko,
H.~Ketterl, W.~Prettl,  and L.~E.~Vorobjev, 
Appl.~Phys.~Lett. {\bf 77}, 3146 (2000).

\bibitem{Vasyukov2010} 
D. A. Vasyukov, A. S. Plaut, and M. Henini, Physica E \textbf{42}, 964 (2010).

\bibitem{Vasyukov2013} 
D. A. Vasyukov, A. S. Plaut, A. H. Macdonald, M. Henini, Int. J. Mod. Phys. B \textbf{23}, 2867 (2009).




\bibitem{kotthaus} A. Lorke, S. Wimmer, B. Jager, J.~P. Kotthaus,
W. Wegscheider, and M. Bichler, Physica (Amsterdam) \textbf{249-251B}, 312 (1998). 



\bibitem{Olbrich2011Ratchet} P. Olbrich, J. Karch, E. L. Ivchenko, J. Kamann, B. M{\"a}rz, M. Fehrenbacher, D. Weiss, and S. D. Ganichev,
Phys. Rev. B \textbf{83}, 165320 (2011).

\bibitem{Olbrich2009Ratchet} P. Olbrich, E.L. Ivchenko, T. Feil, R. Ravash, S. D. Danilov, J. Allerdings, D. Weiss, and S. D. Ganichev, 
Phys. Rev. Lett. \textbf{103}, 090603 (2009).

\bibitem{Kannan2011} 
E. S. Kannan, I. Bisotto, J.-C. Portal, R. Murali, and T. J. Beck, Appl. Phys. Lett. \textbf{98}, 193505 (2011).

\bibitem{IvchenkoRatchet} E. L. Ivchenko and S. D. Ganichev
JETP Lett. \textbf{93}, 673 (2011).


\bibitem{Jiang2011_2}  Ch.~Jiang, Y.~Chen, H.~Ma, J.~Yu, and Y.~Liu,
Appl. Phys. Lett. \textbf{98}, 232116 (2011).

\bibitem{Ganichev2009DMS} S. D. Ganichev, S. A. Tarasenko, V. V. Bel'kov, P. Olbrich, W. Eder, D. R. Yakovlev, V. Kolkovsky, W. Zaleszczyk, G. Karczewski, T. Wojtowicz, and D. Weiss, 
Phys. Rev. Lett. \textbf{102}, 156602 (2009). 

\bibitem{Terentev2011} Ya. V. Terent'ev, C. Zoth, V.V. Bel'kov, P. Olbrich, C. Drexler, V. Lechner, P. Lutz, A.N. Semenov, V. A. Solov'ev, I.V. Sedova, G.V. Klimko, T.A. Komissarova, S.V. Ivanov, and S.D. Ganichev,
Appl. Phys. Lett. \textbf{99}, 072111 (2011).

\bibitem{Olbrich2012} P. Olbrich, C. Zoth, P. Lutz, C. Drexler, V. V. Bel'kov, Ya. V. Terent'ev, 
S. A. Tarasenko, A. N. Semenov, S. V. Ivanov, D. R. Yakovlev, T. Wojtowicz, U. Wurstbauer, D. Schuh, and S.D. Ganichev,
Phys. Rev. B\textbf{ 85}, 085310 (2012).

\end{thebibliography}
\end{document}